# Engineering frictional characteristics of MoS$_2$ structure by tuning thickness and morphology- An atomic, electronic structure, and exciton analysis

Jatin Kashyap*[1], Joseph Torsiello [2][4], Yoshiki Kakehi [3][5], Dibakar Datta[1]

[1] Department of Mechanical and Industrial Engineering, New Jersey Institute of Technology Newark, NJ 07103, USA
[2] Department of Physics, New Jersey Institute of Technology, Newark, New Jersey 07102, USA
[3] Bergen County Technical Schools, Teterboro, NJ 07608, USA
[4] Department of Physics, Temple University, Philadelphia, PA 19122, USA
[5] College of Computing, Georgia Institute of Technology, Atlanta, GA 30332, USA

*Corresponding authors
Jatin Kashyap, Email: jk435@njit.edu

Abstract

2D materials are often used in cutting-edge optoelectronics applications. And that makes it essential to study their mechanical properties. In this study, we performed atomic and electron dynamics analysis to study the impact of morphological and thickness changes of a MoS$_2$ multi-layered system on its tribological properties. We had considered 4 different cases, i.e., number of layers (1-4 layers), number of indents (2-8 indents), the radius of indents (12Å, 16Å, 20Å, 24Å), and pattern of indents (0°, 25°, 30°, 35°, 45°, 60°) resulting into total 18 subcases. We used a MD code to model the movement of a tip over the surface of the MoS$_2$ system. From MD results, we observed that changing the radius and number of indents were the most effective, and changing the number of layers and indents' pattern were the least effective way to tune the frictional characteristics in the MoS$_2$ system. In the ground state ab-initio study phase, analysis shows that as the number and radius of indents increased, the number of stretched bonds in the systems increases. Consequently, the volume covered by the HOMO iso-surface increases, and the volume covered by LUMO decreases. That makes higher area/volume available to lose/share charge carriers, resulting in stronger interlocking between layers and tip. And from TD-DFT calculation, we had observed that not only the excitons were being formed, but they were being formed across the interface of layers and tip, which results in stronger interlocking between the layer's surface and tip despite a decrease in the LUMO iso-surfaces' area/volume. We believe these interlayer excitons result in higher average Z-axis reaction forces for the indents number subcases and lower for indents radius subcases as the number and radius of indents increase. This results in a decrease and increase in frictional forces exhibited for the indent's number and radius cases, respectively since frictional force is a function of the normal reaction of a surface. Further, an in-depth TD-DFT investigation shows that as the number of layers increases, the electrons of a given electron-hole pair delocalize over a bigger area, resulting in weaker interfacial bonds. On the other side, the increase in the layers also increases the excitonic energy, resulting in more robust couplings that counterbalance the weaker interfacial bonds, and neither a significant increase nor a decrease in the frictional force was observed as the number of layers was increased.

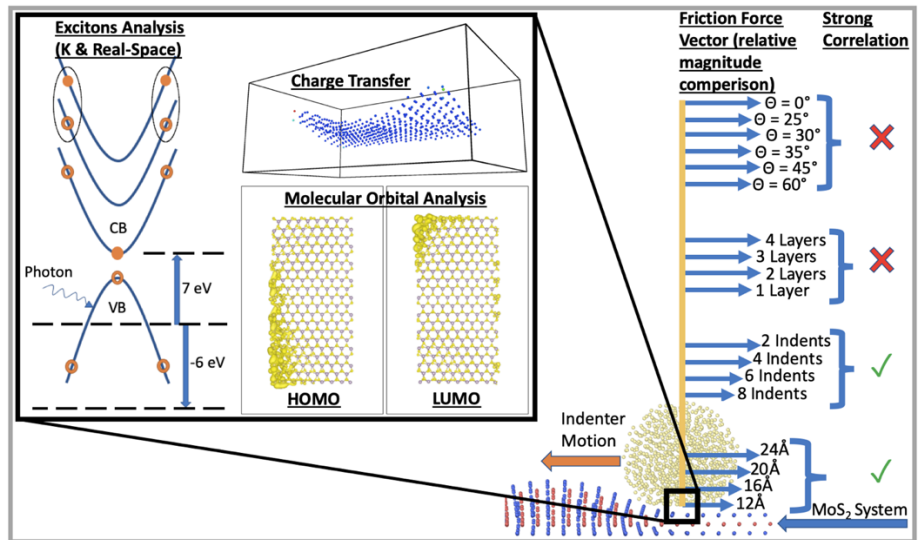

1. Introduction

2D (Two-Dimensional) materials are constituted of a few layers of atoms, and they are known for their cutting-edge applications in the domain of electronics, optics, and magnetics[1]. This work highlights their tribological properties and different approaches to tune them. Whenever a 2D material is used in a device, it will always form physical contact with another body, and consequently there will usually be relative motion across their surfaces. And relative motion between surfaces yield friction. Friction could be desirable for an application, i.e., stopping a moving part, or non-desirable, i.e., supporting a moving part while offering minimum resistance to its motion, ideally none. This study primarily focuses on tuning the frictional characteristics of 2D materials by engineering the surface morphology and thickness[2][3]. While surveying the literature, we had discerned there were mainly two ways by which the tribological properties can be modified of the given layered materials with more importance being given to the former than the latter[4], i.e., a) introducing a structural change in the lattice geometry of material either by wrinkles or indents or folds or doping or functionalization or vacancy generation or straining, etc. and b) non-geometrical changes either by applying electric/magnetic/thermal field or annealing, irradiation, surrounding atmospheric conditions, etc. For tuning the surface morphology of 2D material, a favored mechanism was generating the wrinkles on the surface or other out-of-plane morphological deformations[5]. Since wrinkles usually acquire the shape of the waves, they can be controlled and modeled straightforwardly using the parameters such as amplitude and wavelength. For example, the wrinkles' pattern change on Chemical Vapor Deposition(CVD) grown Graphene had been demonstrated to introduce frictional anisotropy[6].

Multiple experimental studies investigated interfacial strength [7], interface super-lubricity [8] and surface frictional [9] characteristics between Graphene, $MoS_2$, $WS_2$, $WSe_2$, and Atomic Force Microscopy (AFM) tip through Puckering Effect and at high pressure (1GPa) and humidity (51%) among other parameters. Furthermore, Graphene's tribological properties were tuned by fluorination-based surface corrugation in an experimental and MD (Molecular Dynamics) setup [10] and through hydrogenation and oxidation functionalization in an experimental and DFT (Density Functional Theory) framework[11]. Additionally, tribology-based aqueous and oil dispersion characteristics of carbon quantum dots(CQDs) decorated 2D materials (h-BN: $MoS_2$: $MoSe_2$: $WS_2$ and Graphene)[12], and interlayer and surface modification-dependent inorganic layers of Zirconium Phosphate[13] were observed, respectively. In other setups, 2D α-ZrP intercalated with multiple amines exhibited lubrication when dispersed in oil[14], and substitute for ambient conditions solid- and liquid-based lubricants were provided by synthesizing graphene-ionic liquid(IL) based on 1-butyl-3-methylimidazolium iodide, i.e., [BMIM][I] [15]. Moreover, models like spring-block, dashpot, 2D multi-atoms Prandtl-Tomlinson, and micro slip finite element were used to obtain correlations like the dependency of Friction Coefficient on the surface pattern of soft polymer, hierarchical complex and bulk materials [16]–[18], the curvature of multiwall hetero-nanotube complex[19], surface corrugations[20], and wrinkle's wavelength/amplitude in 2D/3D complex[21]. Furthermore, frictional characteristics were controlled by covering SrTiO3's surface with molecular-monolayers[22], redistributing electrons in hexagonal boron-nitride bilayer through intercalation of fluorine atoms[23], and inserting Graphene against Silicon Force Microscopy(SFM) tip [10]. And similarly, another substitute of friction control was generating vacancies (which was more effective than C-O based add atoms functionalization) in Graphene for triboelectric NEMS/MEMS[24][25], and epitaxial Graphene on SiC, which becomes even more effective for bilayer due to electron-phonon coupling[26]. Additionally, Moiré superlattice seems to play a pivotal role in affecting the interfacial or surface frictional characteristics of multiple systems, i.e., Graphene on Ge(111) [27], Graphene flake subjected to area change and biaxial strain [28], fluorographene/$MoS_2$ [29], Graphene on Re(0001) and Pt(111) [30], and Graphene on Ru(0001) and Ir(111)[31]. In several studies multiple correlations regarding tribological characteristics were discovered including dependency of Graphene/Graphene; hBN/hBN; $MoS_2$/$MoS_2$; Graphene/hBN; H-graphene/h-BN system on atom's electronegativity[32], $MoS_2$ system on lattice orientation[33]. And similarly induced in-plane strain caused charge transfer[34], Graphite on $Li^+$ intercalation[35], $MoS_2$; $WS_2$; and $WSe_2$ on thickness[9], multi-layer Re intercalated homo $MoS_2$ on thickness due to enhanced phonon-electron interactions [36].

Another set of experimental/computational studies demonstrated friction tuning in electrochemically reduced graphene oxide(ERGO) can modify supercapacitor's charge storage [37], thickness controlled Zinc acetate (Zn(Ac)$_2$) derived boron nitride nanosheets (BNNSs) can reduce the friction coefficient of pure water[38]. And consonantly friction modification in BN/BN system by perturbing charge transfer caused by changing location of impurity atom[39] and achieving high-temperature lubricity by intercalating 11-aminoundecanoic acid in protonated titanate $H_{1.07}Ti_{1.73}O_4$(organic/inorganic complex)[40]. On the other hand, multiple studies used non-structural parameters as catalysis for affecting frictional characteristics including surface treatment and temperature variations[41], water intercalation and condensation for Graphene, Graphene Oxide, and $SiO_2$[42]. And similarly, ion-irradiation of Graphene and Diamond-like carbon in Nitrogen filled chamber[43], water and alcohol intercalated in Graphene/Mica[44], magnetic field between oxygen molecules layers on Nickel and gold substrate[45], and ambient humidity on the salt surface [46]–[49]. Furthermore, frictional characteristics were affected by the normal electric field on $MoS_2$[28], high pressure (46MPs) annealing of CVD-grown hBN monolayer[50], electronically engineering the ionic liquids between AFM tip and the electrode[51], condensed water from highly humid air between CVD $MoS_2$ and Silica substrate[52]. There were multiple applications utilizing frictional characteristics of layered materials, including quantifying friction exerted by the tip on organic composite[37], 0D/1D/2D materials based friction powered skin-mountable strain sensor[53], friction tuning graphene-oxide(GO)/poly(N-isopropyl-acrylamide)(pNIPAM) composite thermo-responsive hydrogels based Tweezers[54]. And similarly interlayer-friction controlled photoresponsivity of $MoS_2$ based tribotronic-phototransistor[55], $MoS_2$-phototransistor with high ON/OFF ratio and drain-source current[56], interlocking friction in bi-layer $MoS_2$ system governing the fracture/crack propagation[57], black-phosphorous thickness/orientation controlled tribological NEMS[58]. Besides, new models and methods like conventional Prandtl-Tomlinson/DFT [59], electromagnetic field-based quantum chemistry[60], 2D Burridge-Knopoff [61], and nano-scratch and Raman spectroscopy[62] were used to find tribological correlations for 2D electron-hole systems including Fluorographene, $MoS_2$, $WO_2$, and $WSe_2/MoSe_2$. Finally, phosphorene's tribological characteristics demonstrated dependency upon load applied on the tip, sliding direction, pre-strain of phosphorene, adhesion with the substrate, substrate's roughness, and bilayer arrangement[63]. Lastly, elastic conduct of TMDCs materials $MX_2$ (M= W, Mo; X= S, Se) was studied for the first time while modeling tribological properties[64].

In this work, we had modeled the $MoS_2$ multilayered system by considering its nucleus degree of freedom in the Molecular Dynamics(MD) analysis phase, and electron's ground and excited states degree of freedom in the Density Functional Theory(DFT)/Time-Dependent Density Functional Theory (TD-DFT) analysis phase, respectively. We investigated the impact of morphological and thickness changes of a $MoS_2$ multilayered system on its tribological properties. We had considered 4 different cases, i.e., number of layers (1-4 layers), number of indents (2-8 indents), the radius of indents (12Å, 16Å, 20Å, 24Å), and the pattern of indents (0°, 25°, 30°, 35°, 45°, 60°) resulting into total 18 subcases as shown in Figure 1. We used the Large-scale Atomic/Molecular Massively Parallel Simulator (LAMMPS) code to model the movement of an indenter's tip over the surface of the $MoS_2$ system. MD results show that changing the radius and number of indents in a $MoS_2$ system was the most effective way of tuning the frictional characteristics. While changing the number of layers and pattern was the least effective way. The ground-state study phase analysis shows that the number of stretched bonds in the systems increases with the number and radius of indents. This increases the Highest Occupied Molecular Orbital (HOMO) iso-surface volume, and TD-DFT analysis demonstrated that the tip might be responsible for the formation of the excitons in the system since Lowest Unoccupied Molecular Orbital (LUMO)'s volume was decreasing, which justifies the MD phase conclusion. Further, we had performed the in-depth electron-hole pair analysis of an oversimplified number of layer subcase to find the rationale behind the non-significant dependency of friction exhibited.

## 2. System Preparation and Methodology

The simulation cell size along X, Y, and Z directions were 159.581Å, 276.3Å, and 200Å for all cases except for the number of layers, where the Z dimension was: 200Å, 206.498Å, 212.996Å, 219.494Å in layers 1, 2, 3, 4 subcases, respectively. A spherical indenter tip of radius 10Å made from Carbon atoms was developed and relaxed in a separate MD simulation run. Consequently, the relaxed tip was placed 2Å above the top surface of the $MoS_2$ structure. First, the system was relaxed using gradient descent minimization to remove constrained/non-existent bonds. After this step, the tip and $MoS_2$ sheets were treated as rigid bodies for all simulation runs. For $MoS_2$ sheets to be treated as fixed, their atoms were assigned 0 m/s initial velocities, and forces were reset to 0 Kcal/mole-Angstrom at every step of all simulation runs. The next run was of 5000 steps NVT run for an equilibrium of the system at room temperature (300K). During this and subsequent runs, the tip was held with the springs. Two springs were tethered on opposite sides of the tip along the X-axis and centered at the center of the tip with spring stiffness of 10 eV/Å. A similar set of springs were installed along the Y-axis.

Only one spring was installed along Z-axis because the tip needed to be only pushed against the $MoS_2$ surface. The Spring constant of the Z spring was kept 5 eV/Å, lower than the other springs, since we just needed to gently push down the tip against the $MoS_2$ sheet with the least impact on the sheet's interaction with the tip. $MoS_2$ sheet works as a balancing spring, and the interaction between the tip and MoS$_2$ sheet counteracts the forces generated by the spring placed on the other side of the tip. These springs for the last run were kept stationary due to the fixed tip. But they were moved along with the tip for the next run. Once the system was visually inspected to not show any abnormal behavior during this equilibration run, the simulation was moved forward with the next run. The next and final run was a 100k steps NVT run (Figure 2). During this run, the tip was moved along with springs holding it in its place together (Figure 2(D.1-D.5)). The difference of springs from previous runs was the spring constant of X and Y springs was increased from 10 to 18 eV/Å, and there was only one spring along Y-axis used to pull the tip along Y direction with a spring constant of 2 eV/Å, increased from previous 5 eV/Å. During this run, the X, Y, and Z components of the torque and force experienced by the tip were measured (Figure S1-S8). For the force measurement, the force was measured on every atom of the tip, and then it was summed to a scalar value. For torque measurement, LAMMPS's inbuild torque measuring command was used.

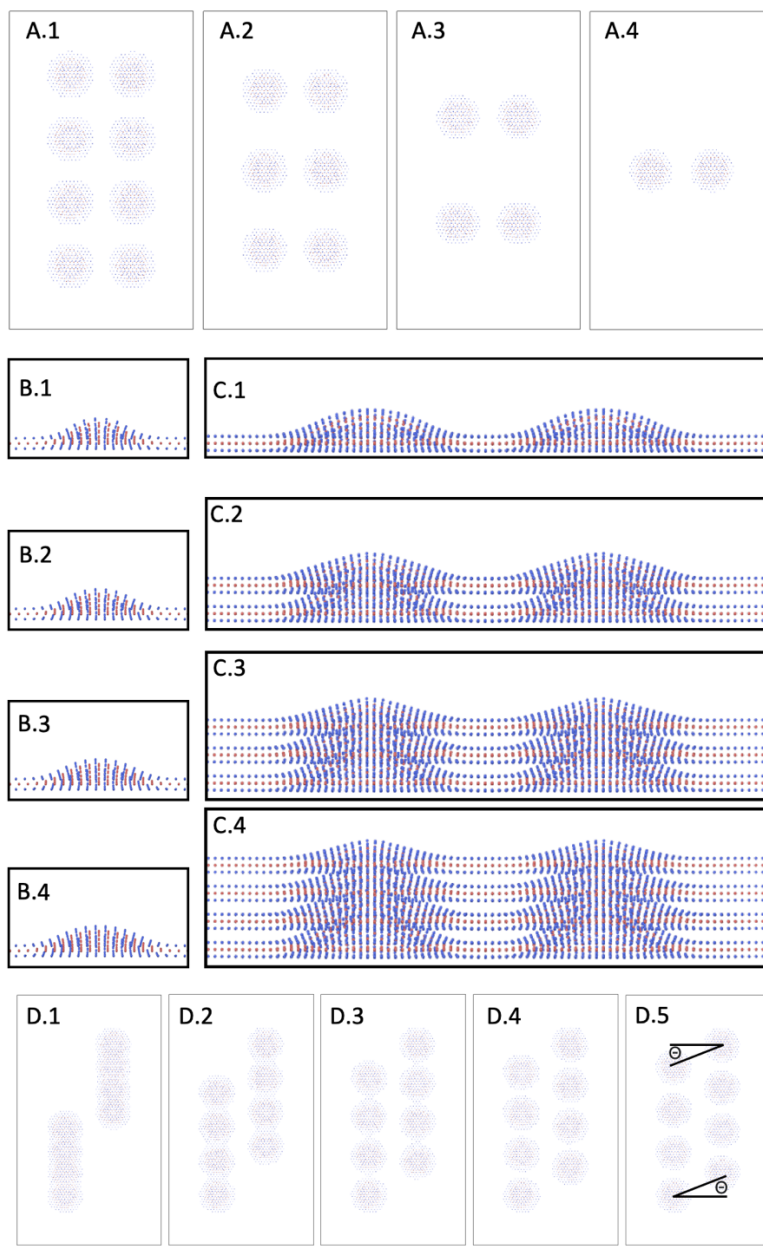

*Figure 1 $MoS_2$ system models considered. A.1-A.4 Shows the change in number of indents (8, 6, 4, 2). B.1-B.4 Shows the change in radius of indents (12 Å, 16 Å, 20 Å, 24Å). C.1-C.4 Shows the change in number of layers (1, 2, 3, 4 layers). D.1-D.4 Shows the change in angles between adjacent indents (Θ=0°, 25°, 30°, 35°, 45°, 60°). For easy visualization, only indents are shown in A and D.*

The main objective of this project was to tune/engineer the forces acting along Y-direction on the tip. Given the morphological features introduced in the MoS$_2$ monolayer, all of them can be attributed to the change in Y-directional force in one way or the other. Since forces are often associated with a moment and hence torque, we had characterized the torques as well. Torque is a rotational effect a linear force had around an axis. In this system, more than one linear force produces the torque. Forces along the X and Z-axes produce the torque around the Y-axis, the Z and Y-axial forces produce torque around the X-axis, and X and Y-axial forces produce the torque around the Z-axis. Due to the interactions occurring at the atomic level, the significant part of the motion of the tip over the MoS$_2$ monolayer was the stick and slip nature. Consequently, we saw the fluctuations in the forces and torques plots in the MD phase of the study (Figure S1-S8). This data noise was filtered out by averaging the forces and torques in the post-processing of the MD phase. Even though these fluctuations were detrimental to the overall data analysis, one can quickly pinpoint an MD step over the whole MD data plot, which can be studied in detail by itself and will help understand broader mechanistic detail about tip's overall motion. A wider perspective had emerged from averaging out similar MD steps. In our case, we had to pinpoint salient MD configurations that we believe had significant contributions to the overall perspective, and then we performed further studies on those configurations using the first principle approaches. The ab-initio codes do not accept large systems considered in the MD phase. Therefore, we had to trim down the system just enough to be fed into DFT code and in a way that it should had maintained salient features of the geometry, if not all, which had contributed significantly to the MD force/torque plots. For all the cases, the number of atoms was kept constant except for the number of layer subcases since, implicitly, the number of atoms will change as the number of layers was changed.

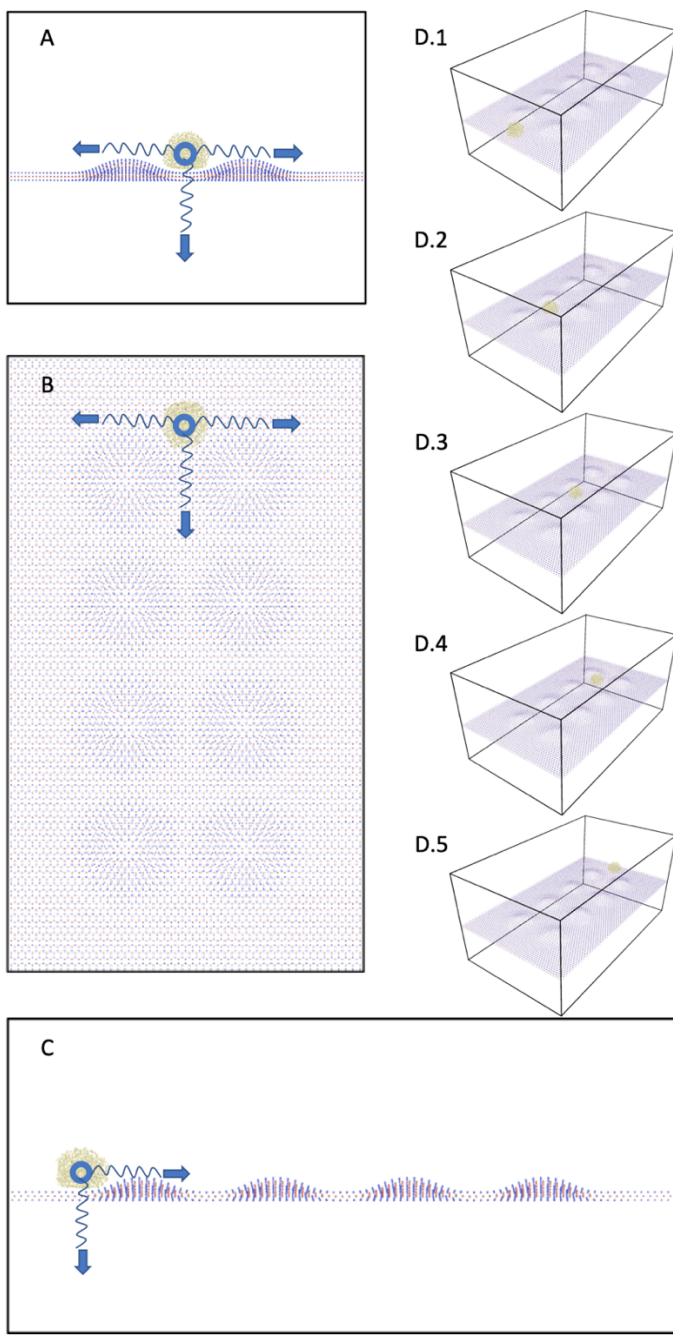

Figure 2 Tip moving across the 1-layer 8-indents (24 Å radius) MoS$_2$ system with Θ=0°. A, B, and C shows the front, top and side view respectively. D.1-D.5 showing one full cycle of the movement of the tip from one end to another end of the MoS$_2$ sheet. Blue arrows are showing the direction of forces exerted by attached springs to the tip.

We had implemented first-principles DFT with plane-wave basis sets and pseudopotentials to describe the electron-ion interactions as implemented in VASP. All calculations were done using the projector augmented wave (PAW) pseudopotentials and the Perdew–Burke–Ernzerhof (PBE) exchange-correlation functional. The plane-wave basis sets were used with a plane-wave cutoff energy of 300 eV. The convergence threshold of Kohn–Sham equations was set to 1$e$–06 eV. The gamma-centered $k$-point sampling grids obtained using the Monkhorst–Pack method were 3 × 3 × 1 with a zero offset for the graphene unit cell. The valence electrons for Mo belong to 4p5s4d orbitals, and S belongs to s2p4 orbitals. The Koopman theorem was used for calculating the HOMO and LUMO iso-surfaces of the systems[65], and the Bader charge program[66] was used to perform the atomic charge transfer analysis. Before DFT calculations, all atoms in the cell and the lattice dimensions and angles were relaxed to the equilibrium configurations using MD. For the band structure calculations, the symbols and coordinates of the high-symmetry points in the first Brillouin zone of the crystals were taken from the MaterialsCloud server[67]. MATLAB and VESTA codes were used for the post-processing of the results. Furthermore, for electron-hole (exciton) calculations, we computed the Bethe-Salpeter Equations (BSE) using **exciting** code[68], which is a full-potential all-electron density-functional-theory package implementing the families of linearized augmented plane-wave methods. Since it's costly to perform such calculations, we must be conservative in selecting the system's size. As one indent in our systems contains 427 atoms and that was too big for BSE analysis, we had to consider only that case where analysis can be done without considering the indents. Given the technical constraints, we can afford to ignore indents only in one case out of four cases considered and still establish a meaningful correlation, i.e., the number of layers case. Therefore, we had considered primitive cell of MoS$_2$, having 6, 12, 18 and 24 atoms for 1, 2, 3, and 4 layers sub-cases alongside appropriate assumptions while explaining the rationale behind the results from the MD phase of the corresponding subcases.

## 3. Results

The net force acting on the diamond tip can be divided into three fundamental forces, i.e., X, Y, and Z-axis. Of these, only the Y-axis force was vital in studying the system's nano-frictional characteristics. But there may be situations where non-Y-axis forces contribute to the Y-axis force, due to which a brief analysis of all the components was needed in this study. Z-axis force is a direct normal reaction from the $MoS_2$ surface due to the diamond tip pressing against it. Any change in the surface morphology will directly impact its magnitude. Similarly, the Y-axis force is the direct reaction to the motion of the tip over the $MoS_2$ surface. Akin to the normal reaction in the case of Z-axis force, the velocity of the tip was maintained constant for all cases and was considered to be a standard reference for comparison purposes. In addition to these primary reasons for the forces to exist, there can be secondary reasons. Those were mainly due to the tip's highly stochastic stick and slip motion over the $MoS_2$ surface. The position of the tip with respect to its relaxed position at the time of the first $MoS_2$ surface morphological change experience by the tip highly influences these secondary reasons because this will set the tone for future movements of the tip and it was highly unpredictable. The secondary rationale for the Z-axis and Y-axis forces appear to had become the primary reason for the average X-axis forces experienced by the tip. That was why we had occasionally observed the average X-axis force, which does not agree with the hypothesis that supports the existence of the average Y-and Z-axis forces.

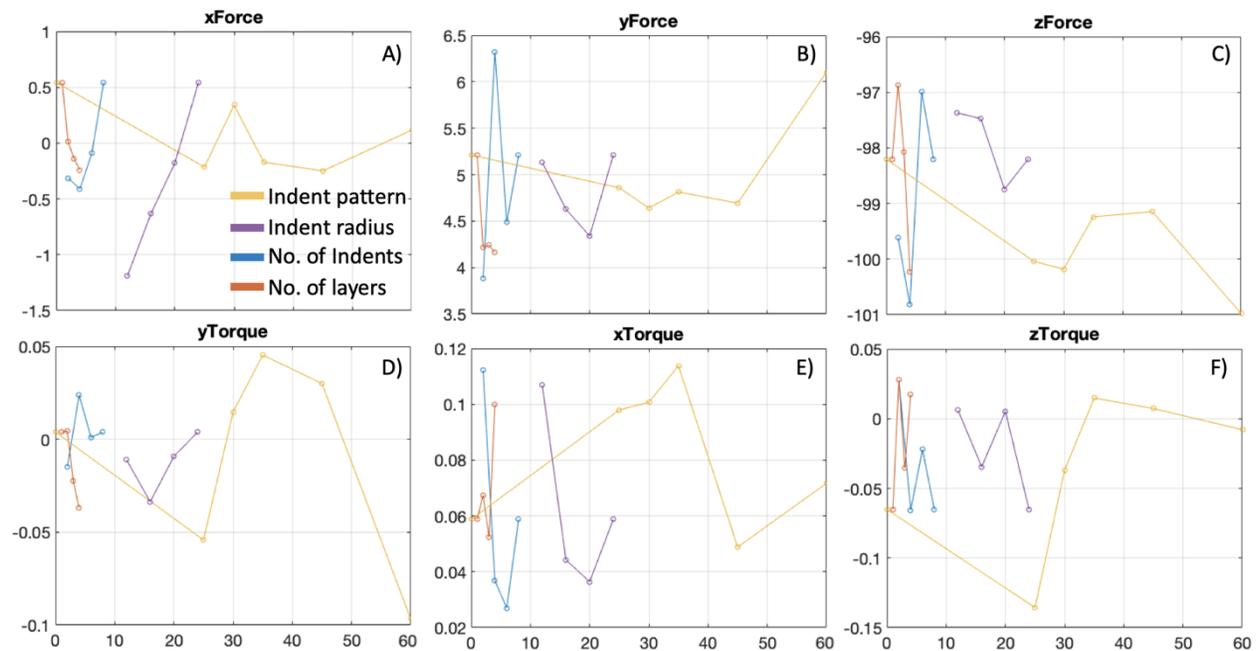

*Figure 3 Average forces along X, Y, Z axes in A), B), C) respectively. Average torque about X, Y, and Z axes in D), E), and F) respectively. Legend is visible on A) for different line colors.*

A plausible rationale for reactional forces to exist could be when the tip had its first interaction with the indent, the tip may had been present on the right-hand side about the centerline parallel to the Y-dimension of $MoS_2$ sheets due to the stick and slip phenomenon. Since it was already shifted to its right-hand side, the interaction of the tip further pushed it down the right-hand side so that it could not recover its initial position throughout the whole cycle of movement over the MoS$_2$ layer. A similar argument can be made for the left-hand side of the tip. On the other hand, the first interaction was probably strong enough to create the over-swinging pendulum motion of the tip, given the spring constant of springs holding the tip in its place. To make the movement of the tip more complicated, the resulting net motion of the tip could be either of these scenarios or any combination of them. But a conclusion that can be made unequivocally is that this will be only possible in the average X-axis force and not for the Y and Z-axes forces. Because the wobbliness of the tip was not "strong enough" to overcome the primary reasons, as outlined above, in the case of X and Y-axes forces. And to avoid the data noise and given the Z-axis force (normal force) had a significant impact on the Y-axis force (frictional force) due to the Law of Friction[69], we shall focus our detailed discussion only on the number of indents and radius cases. Because only these two cases demonstrate significant linear correlation on the Y and Z-axes force plots (Figure 3). Furthermore, an increase in the number and radius of indents results in a higher population of stretched bonds in the whole system, leading to a significant change in the electronic structure and ultimately a corresponding transition in the mechanical behavior of the system. As opposed to the number of layers and indents pattern change subcases, where the changing the number of layer and indents angle doesn't produce a significant switch in the number of stretched bonds and hence the electronic structure in/of the whole system resulting in nonlinear changes on average Y and Z-axes force plots. Furthermore, Figure 3 shows that the first point of the orange curve and yellow curves, and the last points of the blue and magenta curve had the same Y-axis values on all of the MD plots since they were all the same structures, i.e., 1 layer having 8 indents of 24 Å radius with 0° indents' pattern.

3.1 Radius

As stated earlier, our primary focus will be the reaction force on the diamond tip along the Y and Z-axes since friction force (the Y-axis reaction force) is the function of the Z-axis reaction force. As the radius increases (Figure 3B), the Z and Y-axis average reactional forces decrease, considering the 24Å subcase as an outlier. The resultant force of these two forces was acting at an angle between 0° and 90°, opposite to the tip motion and spring pressing down the tip on the MoS$_2$ surface. And as the radius of the indents increases, the reaction forces exerted by the indents on the moving tip also increase since the tip moves with constant velocity among all the radii considered. The higher resultant reaction force was being split into the higher reaction force components along Y and Z-axes, resulting in the increased magnitude of the Y and Z-axes reactional components as the radius increases (Figure 3B). For discussion, we can call this resultant reactional force the sum of the upstream reactional force since it comes into existence when the tip touches the indent for the first time on its face-up and downstream reaction force, which was the force experienced by the tip while leaving the indent. This can be described by the superposition of all the forces in the Y-Z plane as we were not considering the X-axis reactional forces. Since the X-axis reactional forces were more indicative of the initial perturbation of the system when the tip touches the 1$^{st}$ indent of the surface rather than indicative of the reactional forces experienced by the tip throughout its overall trek over the MoS$_2$ sheets. And since these observations were made from force plots obtained by averaging the forces experienced by the tip throughout its complete motion cycle over the MoS$_2$ sheet, the upstream and downstream reactional forces had to add up to satisfy the proportionalities observed.

There could be two reasons that this Y-axis reactional force was emerging. First, as the radius increases, the Y-axis component of the force exerted by the indent on the tip increases. As this force rises, the Z-axis torque exerted on the tip increases. Comparing the Z-torque charts for radius and pattern change-subcases highlights a clear trend in the latter but not the former. This can be attributed to the fact that in the case of pattern-subcase, the indents were placed asymmetrically, and if an indent exerts a torque on the tip, there was no indent on the other side to counteract, except in the inline subcase. However, this was not the situation for radius subcases. An indent on the other side counteracts the torque exerted by force generated by an indent on one side of the tip. On the other hand, the second mechanistic pathway may be the average Y-axis force on the tip could had resulted from the stiffness reaction exerted by the springs holding the tip. As the radius of the indents increases, the negative Y-axis forces exerted on the tip increase. And once the tip gets past the indent, the reaction forces faced by the tip exerted by the springs holding the tip get leveraged and push the other side of the tip by the amount greater than the initial Y-axis force exerted by the indent on the tip. This also results in the same outcome, i.e., an average Y-axis reaction force increase while clear or unclear trend on the Z-torque for indent's pattern case or radius increase case, respectively (Figure 3F).

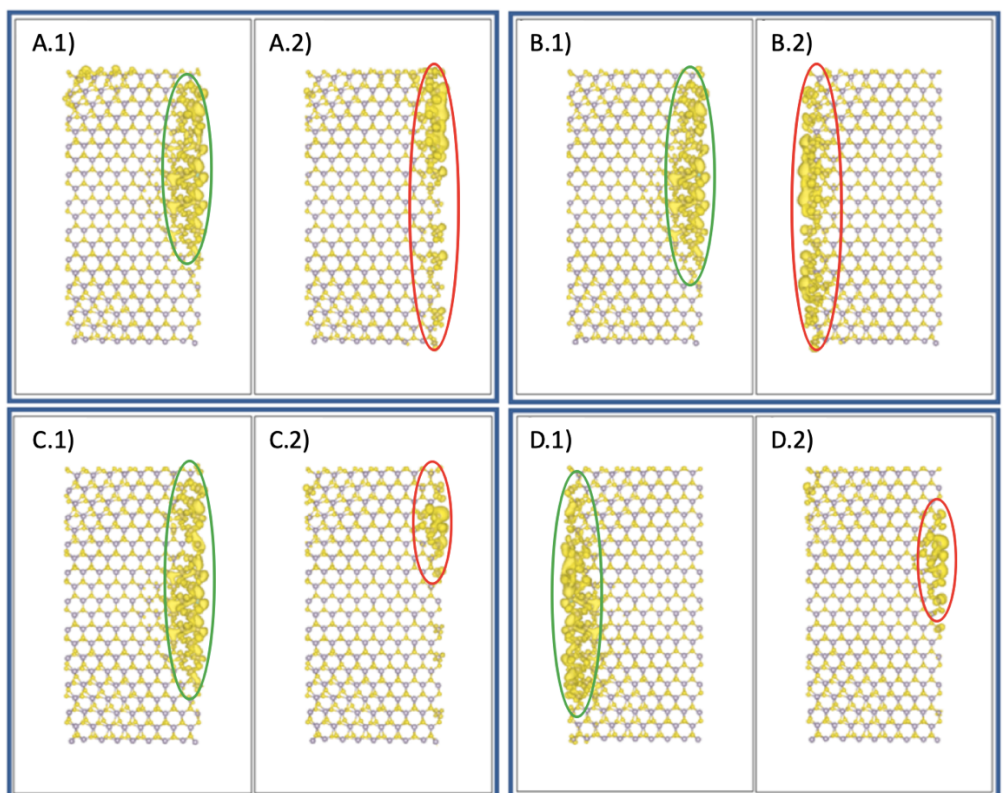

Figure 4 Molecular orbitals of radius sub-case. A), B), C), and D) shows the radius of 12Å, 16Å, 20Å, and 24Å respectively. 1s(left) are HOMO and 2s(right) are LUMO iso-surfaces.

In electronic structure analysis, the volume covered by the HOMO-isosurface increases, and LUMO decreases as the indent's radius increases (Figure 4), which can be attributed to the bond stretches. As the radius of indents increases, more stretched bonds among Mo and S atoms in the system occur. The rise in HOMO volume dictates that the monolayer was willing to lose or share the electrons through a larger area. And the decrease in the LUMO volume dictates that the lesser volume of layers was accepting the electrons. But we already know from the MD force/torque results that as the radius increases, the Y and Z-axes reaction forces also decrease. That means the electronegative parts, i.e., the volume occupied by HOMO of the monolayer, were less likely to lose/share its charge carriers with the other parts of the monolayer, i.e., the volume covered by the LUMO. But since the Y and Z-axes reaction forces were decreasing, the interlocking interaction between the tip and layer increases, i.e., the layer pulls down the tip with higher strength. And this could be only possible if the greater HOMO-covered volume of the monolayer shares its charge carriers with the tip because the tip was the only other body contacting layer surface. And since we cannot model the whole system in ground-state DFT, it's difficult to prove it conclusively. But the Bader charge transfer analysis and the MD force analysis point to this possibility.

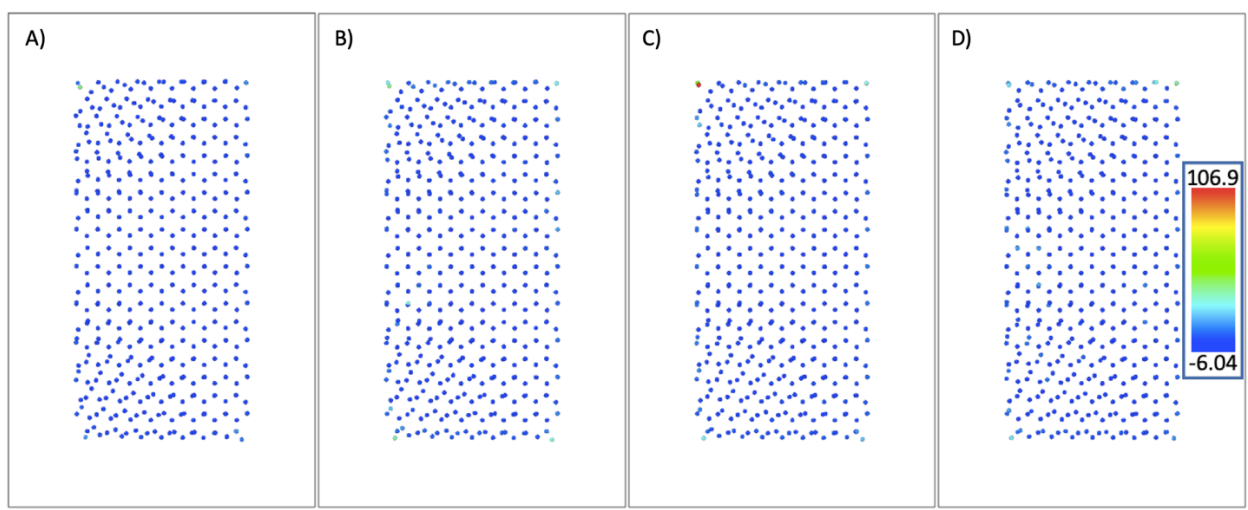

*Figure 5 Bader charge transfer of the radius sub-case. A), B), C), and D) shows the radius of 12Å, 16Å, 20Å, and 24Å respectively. Ligand with its end values is visible in inset of figure D in eV.*

Additionally, the Bader charge transfer analysis (Figure 5) indicates the atoms of this system had a high charge transfer range compared to the range observed in the number of layers and indent pattern cases. The same holds for the number of indents case as well. The ranges were 112.94 eV, 98.97 eV, 34.75 eV, and 1.78 eV for indents' radius, number of indents, number of layers, and indents' pattern subcases. It agrees with the results obtained from HOMO/LUMO plots that due to significant change in the stretched bonds density and hence electronic structure, the radius and number of indents subcases were displaying a strong correlation with the resulting frictional force changes. And this was not true for the number of layers and indents pattern cases because no significant change in the stretched bonds density occurs as the input parameters were changed, which was also visible in their Bader charge transfer plots, i.e., having a lesser charge transfer range. In Bader charge plots (Figure 5), it must be highlighted that the edge states tend to had electropositive behavior, and flat portions as electronegative behavior. And this can be attributed to the highly stretched and/or dangling atomic bonds at edges due to trimming the systems to accommodate the modeling within the available computational resources. Further, we had observed in the exciton calculation part for the multi-layer case (section 3.3.1) that electron-hole pseudo particles (excitons) were being formed between holes in the $MoS_2$ layer and electrons from the tip (Figure 10). This further supported the results observed in this section that as the radius of indents increases, more couplings occur between charge carriers across the layers and the tip, which results in the stronger interlocking between these two parts of the system resulting in a lower average Z and Y-axes reactive forces. To summarize, a change in the surface tribological properties will be pronounced and easier to detect in the reactional forces experienced by the moving tip if the surface and tip were "locked in" rigidly. And this was not observed for the number of layers and pattern change cases because of reasons already stated in section 3.1.

## 3.2 Number of Indents

In this section, we had investigated the frictional force dependency related to the number of indents. An increase in the number of indents in the MoS$_2$ monolayer increases the number of events that contribute to the interaction of the tip and an indent. And as these events' occurrences increase, it skews the average force and torque calculations. Therefore, as building upon our previous argument of the generation of the average Y-axis reactional force, we can extend the opinion supported by the same hypothesis that as the tip was interacting with an indent in the monolayer, it experiences the net negative Y-axis force either as an act of the indent on tip or reaction of the springs holding the tip. Since one interaction between the tip and an indent increases the average negative Y-axis force, there shall be a linear relationship between these two, i.e., as the number of interactions increases between the tip and an indent, the average negative Y-axis reaction force will increase. And that was precisely what was observed on force plots in Figure 3B, as the number of indents was increased in the monolayer MoS$_2$ from 2 to 8 while considering 2 indents subcase as an outlier.

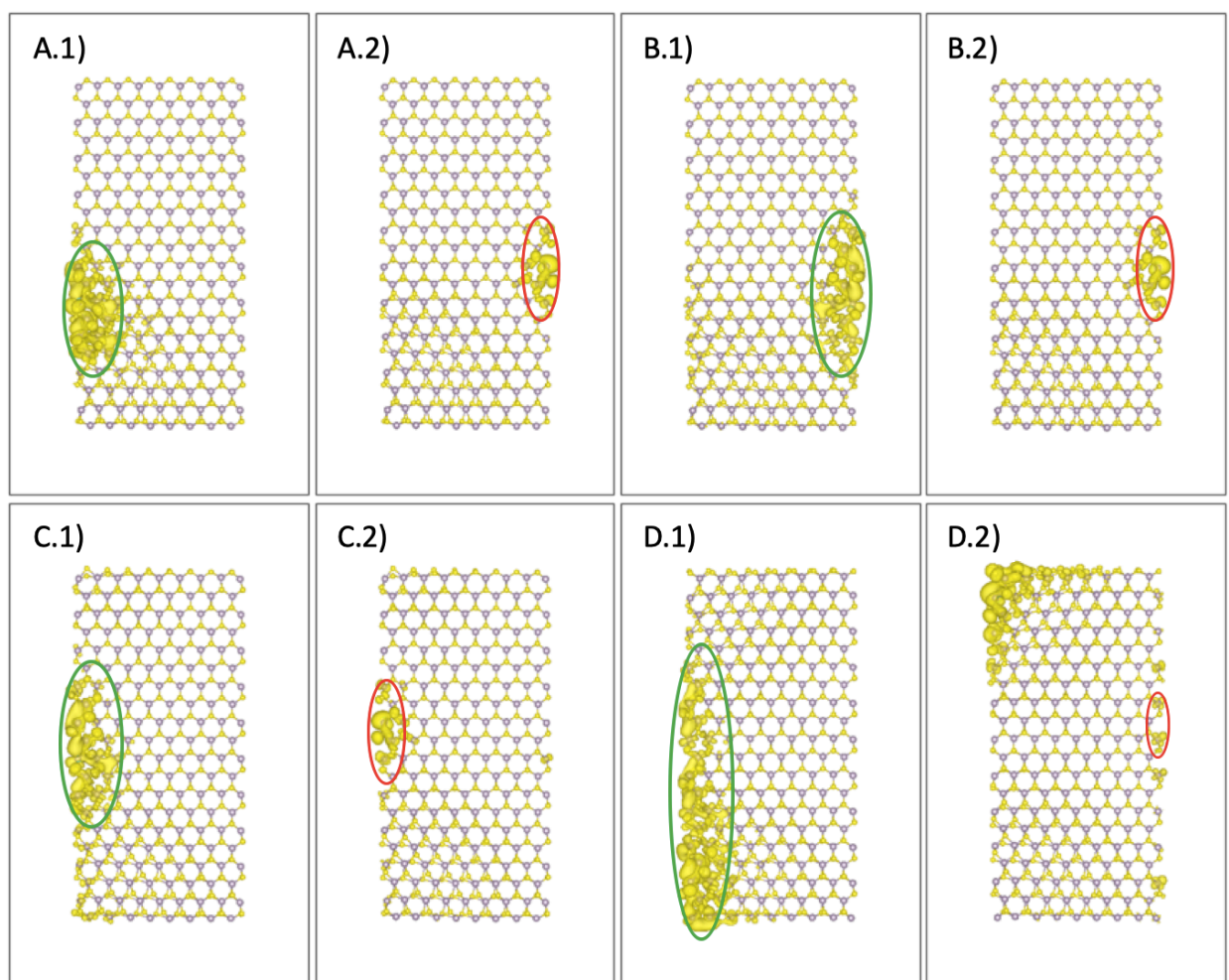

*Figure 6 Molecular orbitals of number of indents sub-case. A), B), C), and D) shows the 2, 4, 6, and 8 indents sub-cases respectively. 1s(left) are HOMO and 2s(right) are LUMO iso-surfaces.*

When the number of indents on the MoS$_2$ varies, the trend in reactional forces was not as straightforward as in the case of the radius subcase. For example, all four data points on the Z-axis reactional force plot were not displaying a line fit which was not observed for Y-axis reactional force (Figure 3B). An almost exact line fit exists if we consider the 2$^{nd}$ data point as an outlier as we considered in the corresponding cases of radius subcase. But given the noise in data because of the highly unstable stick and slip motion of the tip over the MoS$_2$ surface, we can make an exception and cluster the last two and first two data points and consider the directly proportional relationship between the number of indents and Z-axis reactional force experienced by the tip. On the other hand, one can observe the increase in the Y-axis reactional force on the tip as the number of indents were increased, considering the 2$^{nd}$ data point as an outlier. To summarize, we observed the increase in the indents was directly proportional to Z and Y-axes reactional forces. The resultant reaction force created by these two components of the reaction force experienced by the tip points along the positive Y-axis and negative Z-axis. Resulting in pushing the tip along its direction of motion while dragging it into the MoS$_2$ surface. As explained in the radius subcase section about upstream and downstream reaction force, in this case as the number of indents increases, the downstream reaction force acting along the positive Y and Z-axis also increases and by greater magnitude in comparison to the upstream reaction forces.

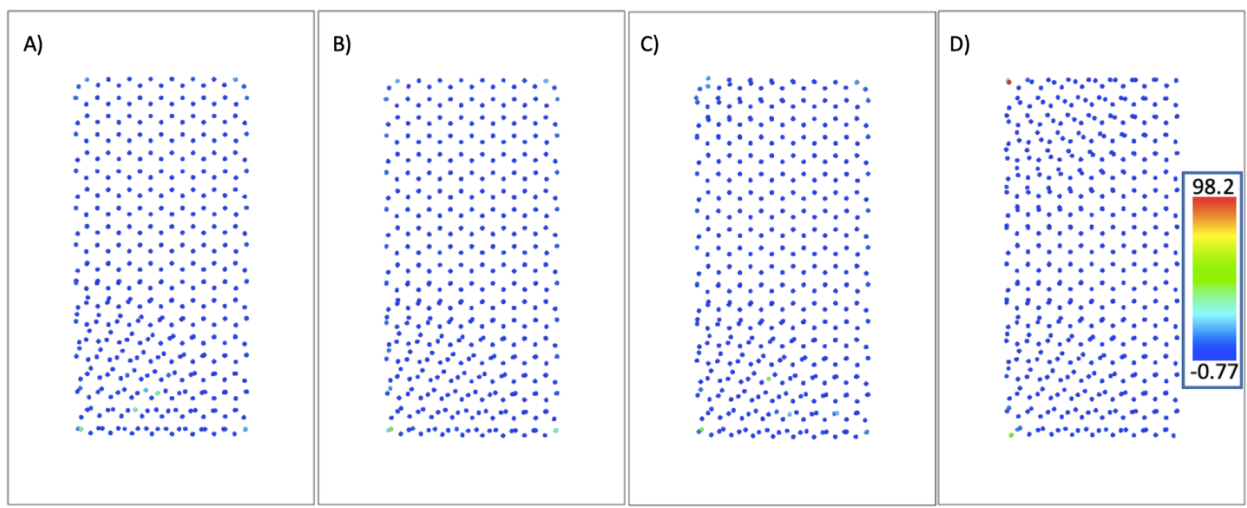

*Figure 7 Bader charge transfer of the radius sub-case. A), B), C), and D) shows the 2, 4, 6, and 8 indents sub-cases respectively. Ligand with its end values is visible in inset of figure D in eV).*

Additionally, the same phenomenon as of indent's radius case, was observed in electronic structure calculations. As the number of indents in the layers increases, the number of stretched bonds increases in addition to increasing the stretchiness of already existing and new bonds. This results in higher volume covered by HOMO and lower volume covered by LUMO-isosurfaces (Figure 6) highlighted in green and red curves, respectively. A decrease in LUMO iso-surfaces dictates the intra-layer(s) bond formation was being deterred as the number of indents was increased. It means more layer surfaces area was available to share the charge carriers with a separate structure, i.e., tip in this case. But as described in the indent's radius case, the bonded interaction among electron-hole pseudo particles occurs as observed from excited-state calculations performed in the multi-layer subcase (Figure 10). And these bonded interactions use the higher HOMO volumes to form a stronger interlocking interaction among the layer and tip that results in a more "rigid transfer" of the changes in the electronic structure to the mechanical behavior, i.e., a higher number of indents leads to lower Y and Z-axes average reaction forces, respectively. Consequently, increasing the frictional force exhibited by the layered system. Similarly, the high range of Bader charge transfer was observed, i.e., ~100 eV in this subcase on par with the indent radius subcase (Figure 7). This again agrees with our hypothesis that the number of stretched bonds was increasing in a similar trend as of indent's radius case, which was much higher than the number of layer and indent pattern cases.

3.3 Number of Layers

For number of layers case, we had observed that the average Y-axis reaction force was almost constant (Figure 3B). This can be attributed to two reasons: first, no force was experienced by the tip from the substrate, and second, the tip was experiencing forces but immediately, an opposite force counterbalances it. It gives the illusion that the tip was experiencing no reactional force. We believe in the second rationale, which can be attributed to non-uniform X and Z-axes force curves in Figure 3B. If it was the former reason, i.e., the tip was not experiencing any force, the probability of average X and Z-axes forces having non-constant nature as we change the number of layers shall be acute. As the number of layers increases, the van der Waals interaction between the multi-layer system increases. We postulate that this was responsible for amplifying the average X, Y, and Z-axes forces exhibited by the substrate on the tip as the number of layers increases. Furthermore, increasing the number of layers increases the substrate's structural stability, augmenting the reaction forces experienced by the tip moving across the substrate/layers. As the number of layers increases, the tip's adherence to the morphological change shall become more rigid. The Y-axis force experienced by the tip while ascending an indent was almost the same as the force experienced while descending an indent but of precisely the opposite sign. There was nearly zero "hysteresis loss". And irrespective of number of indents the tip was moving over, it will always experience opposing forces along the Y-axis while ascending and descending the indent. That will result in almost zero accumulation of the net average Y-axis reaction force, supported by the Y-axis reaction plot showing the nearly horizontal line. On the other hand, the average Y and Z-axes reactional forces were influenced by the increase in pull-down generated by the rise in the number of layers in the substrate. As the number of layers increases, the intensity of van der Waals interaction between substrate and tip also increases, which decreases the distance between the substrate, which now had a higher number of layers, and the tip. As the substrate cannot move towards the tip since it was fixed, the tip had to move closer to the substrate given an opportunity. And that opportunity was available whenever the equilibrium breaks due to the movement of the tip. In other words, the opportunity arises throughout the whole tip's movement period. The same argument can be easily made for the average X-axis reactional force. In this case, the only difference was that due to the tip's wobbling motion, the tip drifted apart on such a side of the centerline so that the resulting reaction force on the tip was a net negative X-axis force. Which kept on increasing as the number of layers increased due to an increase in van der Waal interaction.

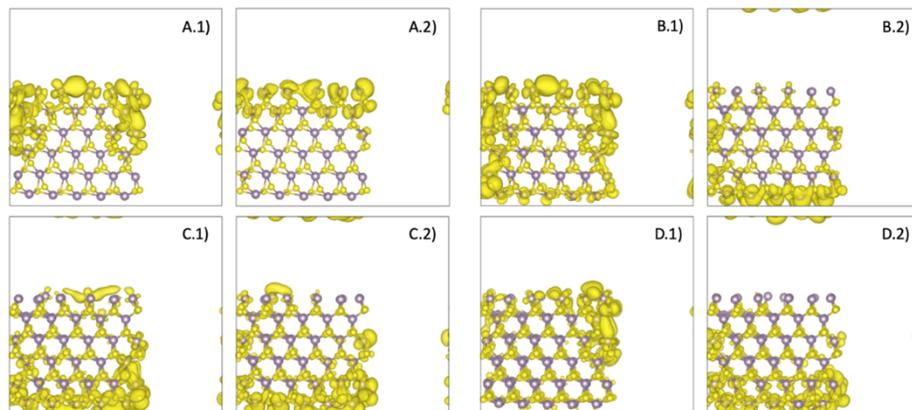

Figure 8 Molecular orbitals of number of number of layers sub-case. A), B), C), and D) shows the 1, 2, 3, and 4 layers sub-cases respectively. 1s(left) are HOMO and 2s(right) are LUMO iso-surfaces.

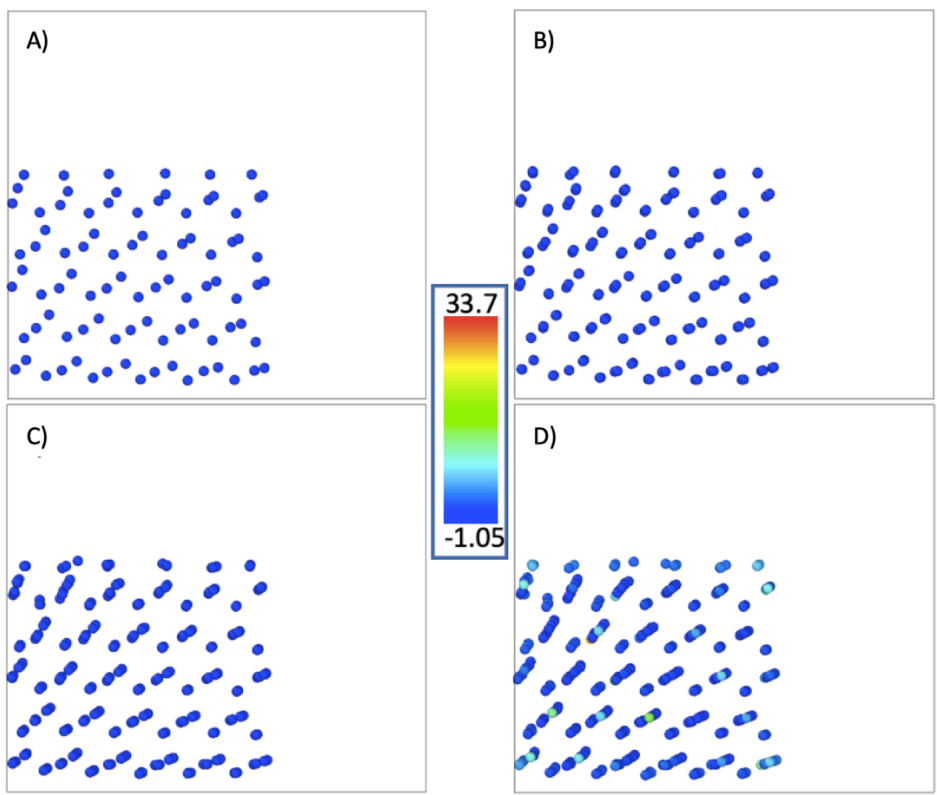

Figure 9 Bader charge transfer of the number of layers sub-case. A), B), C), and D) shows the 1, 2, 3, and 4 layers sub-cases respectively. Ligand with its end values is visible in center in eV.

Not only van der Waals interaction but bonded interaction can also demonstrate a significant role, as per HOMO/LUMO orbital plots and Bader charge analysis in Figure 8 and Figure 9, respectively. In this case, the FMO plot was not as helpful as the Bader charge plot. As the number of layers increases, the atoms with the less dark blue color appear, which indicates the losing/sharing of the electrons and hence the probability of bond formation. On the other hand, there was no clear trend observable on X and Z-axes torque (Figgure3D-3F), but there was a coherent decreasing trend on Y-axis torque (Figure 3E) as the number of layers were increased. Y-axis torque was produced by more than one force acting in the XZ plane (Figure 3E). Average X and Z-axes forces show an inverse trend with the number of layers. That results in these two forces trying to counterbalance the moment generated by each other since they were working in the same plane and around the same axis over the same range of the number of layers. Since the decrease in average Z-axis force was almost twice as big as the decrease in the average X-axis force, the moment generated by the Z-axis force was bigger, resulting in the net negative torque around Y-axis as the number of layers was increased. Furthermore, as explained in the previous two sections, the range of the Bader charge transfer was comparatively less than the last two subcases, i.e., 34.75 eV. This further supported our argument that the strong bonding between the surface and the tip accompanied by counterbalance events was the rationale behind why we cannot observe a significant correlation on the Y-axis forces plots despite moderate changes in the electronic structure of the $MoS_2$ system for this subcase.

3.3.1 Quasi Particles in multi-layers case - Excitons

We performed the excited-state calculation using the Bethe-Salpeter equation (BSE) model to explain the MD phase results further. Despite already trimming down the systems from MD step into DFT step analysis, the structure was cut further and over-simplified since the cost of the computations increases exponentially for excited states calculations with the BSE model. Consequently, we had only consider number of layers case subsystem, which demonstrated non-significant frictional characteristics. We had skipped the remaining 3 systems of pattern, radius, and the number of indents due to our inability to converge the systems within available computational resources.

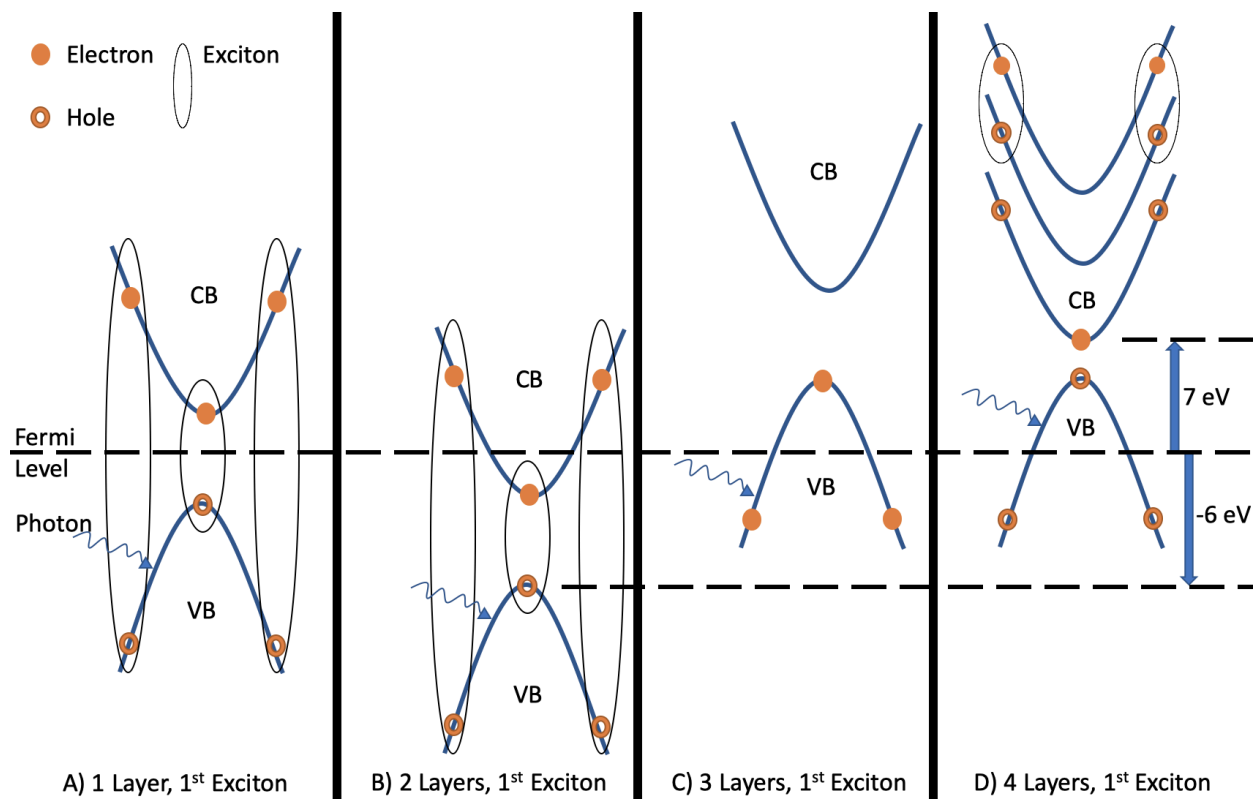

*Figure 8 Schematic exciton visualization in k-space. A), B), C) and D) shows the excitons for the 1, 2, 3, and 4 layers subcases.*

Excitons are electron-hole pairs that usually act as primary drivers for 2D materials' properties. Since exciton wavefunctions are 6-dimensional matrix objects, we need to reduce the dimensionality to handle the visualization by keeping one object fixed and plotting the cloud of another object in electron-hole couples. In this work, hole location was kept fixed while plotting the cloud density of the electron distribution for a given exciton. For one-layer system, hole located at x=0.02, y=0.02, z=0.5 scaled coordinates were fixed in the vicinity of a Mo atom located at x=0.0, y=0.0, z=0.5. An offset was provided to avoid the spurious effect generated in the real space exciton visualization due to the exact overlapping of the hole of a given exciton and an atom (Figure S10). To provide the comparative analysis for the higher number of layers subcase, we need to keep a reference point to account for which the identity of the fixed hole was kept the same.

As Figure 10 and Figure S9 show, the excitons' energy increases as the number of layers increases if we consider the 1-layer (1L) case as an outlier. Furthermore, we observed from the fatband plots (Figure S9) that the coupling coefficients between electron-hole pairs decrease as the number of layers increases. But we shall be careful about fatband plots since they were for highly simplified versions of the number of layers subcases. One of the major simplifications was that they include neither the tip nor the indents in the modeled system. Given that, we had attributed the non-significant change in frictional forces as the function of the number of layers to the two contributing factors mainly, i.e., the electron pair based covalent bond and electron-hole coupling. From the fatband plot, we can observe that the electron-hole coupling had an inverse relation with the number of layers in the case of a system containing only layers. But when the tip was introduced over the layers, then three outcomes were possible: as the number of layers was increased, i) the electron pair based covalent bonds and electron-hole couplings get strong among layers and tip, ii) the electron pair based covalent bonds gets weaker but the electron-hole coupling was getting stronger by greater magnitude among layers and tip and iii) the electron pair based covalent bonds was getting stronger but the electron-hole coupling was getting weaker by greater magnitude among layers and tip. And finally, the greater number of coupled particles and stronger coupling among these particles produces greater interaction between tip and layers, contributing to higher frictional characteristics as the number of layers was increased. From the Fatbands (Figure S9), delocalization of excitons over the K-space was observed, implying that they must be localized in real space (Figure S10). Real-space visualization (Figure S10) shows the electron clouds getting stretched along the system's height as the number of layers increases. For the 1-layer system, the electron cloud was localized over the Mo atom and less stretched in the Z dimension but as the number of layers was increased the electron cloud starts getting stretched along the Z dimension and starts covering more of the S atoms of $MoS_2$ sheets and less of Mo atoms. As demonstrated in Figure 10, all of the excitons observed preferred intra-valley formations.

Increasing the number of layers decreases the excitonic effect (the interaction between electron and hole pair of an exciton). The excitonic effect dictates how strongly an electron and hole in a given exciton were paired. As the number of layers increases, the number of atoms also increases, making a bigger crystal size available for an electron cloud to spread over for a fixed hole in a given exciton. As the electron cloud for a given fixed hole in an exciton spreads over a bigger volume, it loses its binding strength with the attached hole than the electron cloud spread in a lesser crystal volume. It may hold true for all the excitons of the system instead of only $1^{st}$ exciton. As the tip was inserted in the system, we believe either no new excitons were being formed, or if they were being created, their generation rate was equivalent to their recombination rate. As discussed above, increasing the number of layers in the system increases the exciton energy. And it was comparatively accessible for an exciton of the layers to couple with the conduction band electron of the tip to form a negative trion or couple with a valence band hole to form a positive trion. Given a trion is a quasi-particle with three charged particles, it had a net tendency to hold components of a system together if the constituting electrons and holes belong to different components. That was why we can see a decrease in the average Z-axial reaction force plot from MD results as the number of layers increases in the system. That supports our hypothesis that the multi-layer system holds onto the tip more robustly as the number of layers increases. But the primary reaction force of this study, i.e., friction force, which was characterized by the Y-axis reaction force on the tip, was almost kept steady as seen from the Y-axis reaction force plot (Figure 3). This can be explained by the wobbliness motion of the tip that regardless of the number of layers the tip was stacked against, there will always be a balancing force against every negative Y-axis movement of the tip due to stick and slip motion. To summarize, we can see an increase in the Z-axis reaction forces felt by the tip traveling across a substrate due to the formation of the charge transfer exciton/trions but derivative of the Z-axis reactional forces, i.e., Y-axis reaction forces OR frictional force experienced by the tip moving across substrate will not be modified. And later overcome the former because of the stick and slip motion of the tip given the cavity spaces between atoms forming the tip and even bigger cavity spaces between the atoms forming the $MoS_2$ layer(s).

3.4 Pattern of Indents:

In the case of pattern change, we investigated how the angle between the centerline of the substrate bisecting X dimension of the $MoS_2$ system and the center of the tip (Figure 1.D.5) affects the frictional force. Only two angles of 0° and 25° were considered for DFT calculations due to computational cost while considering all angles (0°, 25°, 30°, 35°, 45°, 60°) for MD analysis. To explain frictional dependency on indents' pattern, we believe as the density/congestion of the atoms increases under the tip, the tip follows any morphological changes in the substrate more swiftly. Or in other words, the link between the substrate and the tip becomes rigorous. Such was the reasoning in the number of layers case. As the density of atoms increases under the tip by any technique, the volume of the HOMO orbital becomes more distributed between Mo and S atoms (Figure 11). And that helps form the bonded/non-bonded interaction between the substrate and tip. Out of four cases considered in this work, we had observed such behavior only in two, i.e., the number of layers and pattern of indents. That means there was a critical value of density of atoms, below which the rigidity of bonding between substrate and tip was not rigid enough to "lock" the substrate with the tip. Studying the value of that critical volume of atoms is beyond the scope of this article, but we want to acknowledge that such a threshold shall exist. And in this case, the rigidity of the bonding may be so strong that we had observed almost constant average forces along X, Y, and Z-axes. Which was only the Y-axis average force in the subcase of the number of layers increase due to lower rigidity compared to the pattern change case. As the pattern angle was increased, the overlap between the indents increases. This results in a higher density of atoms than usual under the tip, which makes the volume of the HOMO orbital moderately bigger and intricately intervened between Mo and S atoms than the one obtained in the highest number of layers, i.e., 4 layers, subcase. And that was attributed to the constant value of not only the average Y-axis reaction force (as in the case of the number of layers subcase) but also X and Z-axes reaction forces.

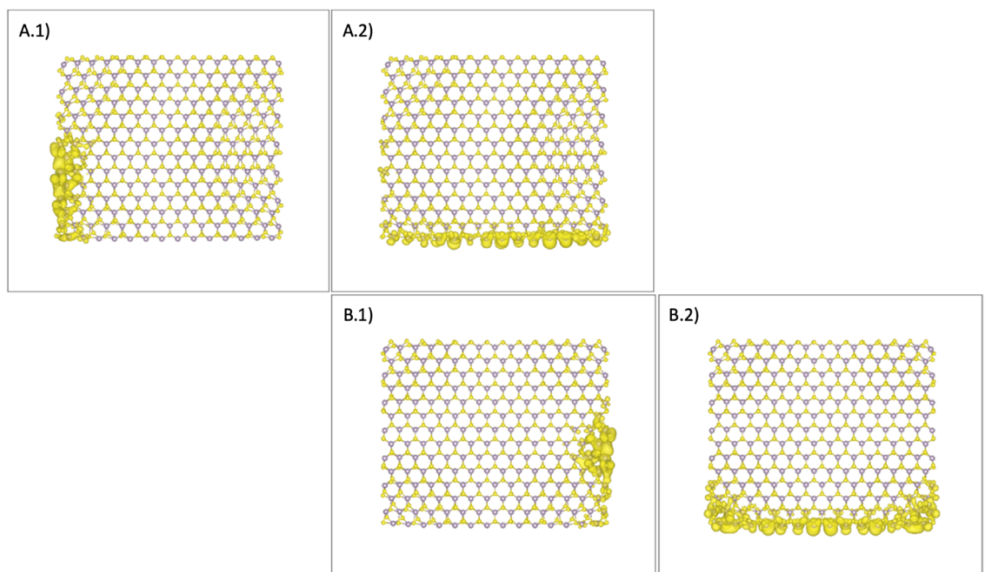

*Figure 9 Molecular orbitals of indents' pattern sub-case. A), and B) shows the Θ=0°, 25° sub-cases respectively. 1s(left) are HOMO and 2s(right) are LUMO iso-surfaces.*

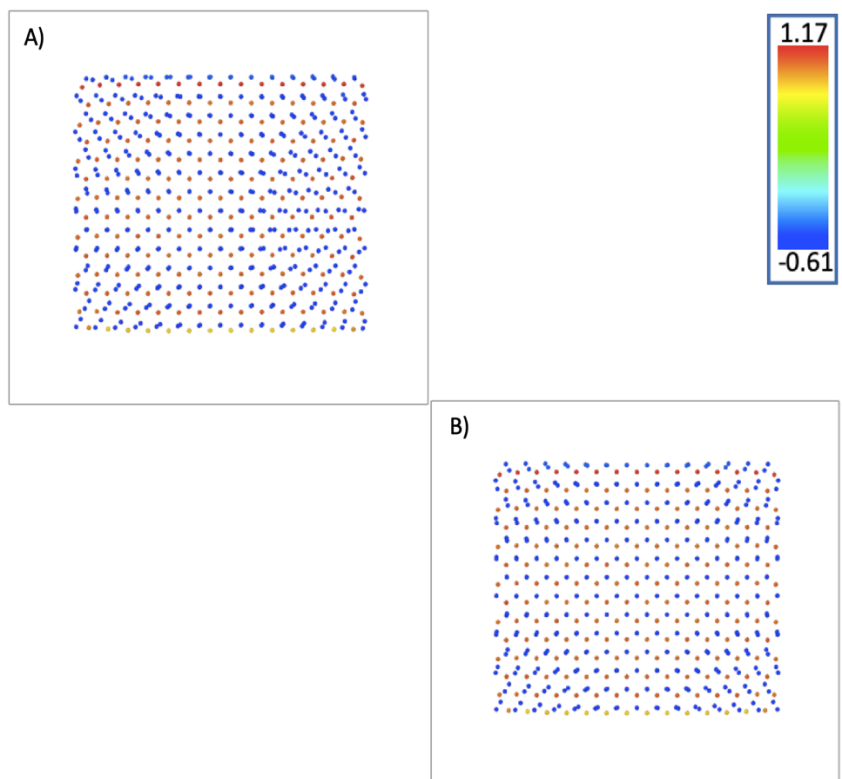

*Figure 10 Bader charge transfer of indents' pattern sub-case. A), and B) shows the Θ=0°, 25° sub-cases respectively. Ligand with its end values is visible in right top*

We observed weak correlations on X, Y, and Z-axes torques, but since forces are what causes the moments and hence torques, and there was an almost constant correlation for all three reaction forces, the probability of formation of correlations in the moments charts was significantly less. That makes a sturdy rationale that the weak patterns observed in the torque charts were not due to the correlations in the force data but were due to the data's noise due to the tip's wobbling motion. As the angle increases, the HOMO iso-surface volume diffused more evenly (Figure 11), and the probability of bond formation improves, which results in the counter movement for every active movement by the tip on the substrate. If there were negative Y-axis forces on the tip while ascending an indent, the equal but opposite force shall also present while descending the same indent. This suggests "hysteresis loss" was reducing as the pattern angle increases. The same argument can be made for forces along the X and Z-axes. This results in the average X, Y, and Z-axes forces and their respective moments/torques being constant. On the contrary, if we consider the range of Bader charge analysis, i.e., 1.78 eV (Figure 12), which was around 1-33% of the significant correlation subcases, i.e., radius and number of indents respectively, it was equally justifiable to consider another rationale. That states the tip was displaying non-significant Y and Z-axes force correlation with the input parameters because the weak interlocking between the layer and tip instead of strong interlocking subjected to a counterbalancing movement for any given reaction along Y and Z-axes resulting in almost flat correlation on Y and Z-axes force plots or noisy correlation at best. With the Bader charge transfer analysis and the MD-based force plots data, our belief was skewed towards later rationale instead of the former, i.e., the tip was in weak interaction with the layers system.

Conclusions

In this study, we had performed atomic and electronic dynamics analysis to study the impact of morphological and thickness changes of a $MoS_2$ layered system on its tribological properties. We had considered 4 different cases, i.e., number of layers (1-4 layers), number of indents (2-8 indents), the radius of indents (12Å, 16Å, 20Å, 24Å), and pattern of indents (0°, 25°, 30°, 35°, 45°, 60°) resulting into 18 subcases in total. We used the LAMMPS MD code to model the movement of a tip over the surface of the $MoS_2$ system. The tip route was chosen in such a way to cover most of, if not all, the critical points covering the structural changes that appear as different systems were considered for analysis. Using the in-build algorithms in LAMMPS, we obtained torque and force experienced by the moving tip about and along X, Y and Z-axes, respectively. We had mainly focused on the Y-axis component of force experienced by the tip, either directly or indirectly. Since the highly stochastic stick and slip nature of the tip's motion results in significant noise in the data, we had averaged the data across the whole production MD run for each of the 18 subcases. It were these averaged values that we had based upon our complete analysis. From MD results, we observed that changing the radius and number of indents in a $MoS_2$ was the most effective way of tuning the frictional characteristics. $MoS_2$ layer tends to offer higher friction to a moving object along the armchair direction as the radius of indents increases and tends to provide lower friction as the number of indents increases. While changing the number of layers and pattern was the least effective way.

Additionally, we had performed ab-initio (VASP) based Bader charge transfer and molecular orbital analysis (HOMO LUMO) to find the rationale behind the results obtained in the molecular dynamics phase. Since the system's size varies from 15,859 atoms (smallest systems-1layer) to 60,859 atoms (biggest system-4layers), even the smallest system was too big to be analyzed in ab-initio code, the structures need to be trimmed. In the ground-state study phase, our analysis shows that as the number and radius of indents increase, the number of stretched bonds in the systems increases. Consequently, the volume covered by the HOMO iso-surface increases, and that of LUMO decreases. That makes higher area/volume available to lose/share charge carriers, resulting in stronger interlocking between layers and tip. And from the BSE calculation, we had observed that not only the excitons were being formed, but they were being formed across the interface of layers' surface and tip, i.e., interfacial excitons, which results in stronger interlocking despite the decrease in the LUMO iso-surface's volume. We believe these interlayer excitons result in higher average Z-axis reaction forces for the indents subcase and lower for the indents radius subcase as the number of indents and indents' radius increase. It further results in a decrease and increase in frictional forces exhibited for the number and radius of indents cases, respectively, since frictional force is the function of a surface's normal reaction (Z-axis). In the last phase of the study, we performed excited state analysis using BSE for an over-simplified replica of the number of layer subcases due to higher computational cost otherwise. As the number of layers increases, the electrons of a given electron-hole pair were delocalizing over a larger area resulting in weaker interfacial bonds. On the other side, the increase in the layers also increases the excitonic energy, resulting in stronger couplings that counterbalance the weaker interfacial bonds, and neither a significant increase nor a major decrease in the frictional force was observed as the number of layers was increased.

**Supporting Information**
Additional information regarding raw data (nuclei and excitonic) for all 18 cases considered.

Figures: Force plots for radius subcase; Torque plots for radius subcase; Force plots for indents' pattern subcase; Torque plots for indents' pattern subcase; Force plots for number of layers subcase; Torque plots for number of layers subcase; Force plots for number of indents subcase; Torque plots for number of indents subcase; Fatbands structures of $1^{st}$ excitons, Real space visualization of $1^{st}$ excitons

**Author Information - Corresponding Authors**
Jatin Kashyap – JK435@NJIT.EDU

**CRediT authorship contribution statement**
**Jatin Kashyap:** Conceptualization, Data curation (DFT/TD-DFT), Investigation, Methodology, Software, Writing – original draft. **Joseph Torsiello**: Conceptualization, Structures generation. **Yoshiki Kakehi**: Conceptualization, Data curation (MD). **Dibakar Datta:** Funding acquisition, Supervision, Validation, Visualization, Writing – review & editing.

**Declaration of Competing Interest**

The authors declare that they have no known competing financial interests or personal relationships that could have appeared to influence the work reported in this paper.

**Acknowledgement**

The authors acknowledge the NJIT faculty start-up grant and the Extreme Science and Engineering Discovery Environment (XSEDE) for the computational facilities (Award Number – DMR180013).

**Data Availability**

The data reported in this paper is available from the corresponding author upon reasonable request.

**Code Availability**

The pre-and post-processing codes used in this paper are available from the corresponding author upon reasonable request. However, restrictions apply to the availability of the simulation codes, which were used under license for this study.

**Bibliography**

[1]   J. Tan, S. Li, B. Liu, and H.-M. Cheng, "Structure, Preparation, and Applications of 2D Material-Based Metal–Semiconductor Heterostructures," *Small Struct.*, vol. 2, no. 1, p. 2000093, Jan. 2021.

[2]   K. Gupta, T. Mukhopadhyay, L. Roy, and S. Dey, "High-velocity ballistics of twisted bilayer graphene under stochastic disorder," Apr. 2022.

[3]   K. K. Gupta, T. Mukhopadhyay, A. Roy, L. Roy, and S. Dey, "Sparse machine learning assisted deep computational insights on the mechanical properties of graphene with intrinsic defects and doping," *J. Phys. Chem. Solids*, vol. 155, p. 110111, 2021.

[4]   S. Zhang, T. Ma, A. Erdemir, and Q. Li, "Tribology of two-dimensional materials: From mechanisms to modulating strategies," *Mater. Today*, vol. 26, pp. 67–86, 2019.

[5]   J. Kashyap, E.-H. Yang, and D. Datta, "Computational study of the water-driven graphene wrinkle life-cycle towards applications in flexible electronics," *Sci. Rep.*, vol. 10, no. 1, p. 1648, 2020.

[6]   F. Long, P. Yasaei, W. Yao, A. Salehi-Khojin, and R. Shahbazian-Yassar, "Anisotropic Friction of Wrinkled Graphene Grown by Chemical Vapor Deposition," *ACS Appl. Mater. Interfaces*, vol. 9, no. 24, pp. 20922–20927, Jun. 2017.

[7]   M. B. Elinski, B. D. Menard, Z. Liu, and J. D. Batteas, "Adhesion and Friction at Graphene/Self-Assembled Monolayer Interfaces Investigated by Atomic Force Microscopy," *J. Phys. Chem. C*, vol. 121, no. 10, pp. 5635–5641, Mar. 2017.

[8]   S.-W. Liu *et al.*, "Robust microscale superlubricity under high contact pressure enabled by

graphene-coated microsphere," *Nat. Commun.*, vol. 8, no. 1, p. 14029, 2017.

[9] L. Fang, D.-M. Liu, Y. Guo, Z.-M. Liao, J.-B. Luo, and S.-Z. Wen, "Thickness dependent friction on few-layer MoS2, WS2, and WSe2," *Nanotechnology*, vol. 28, no. 24, p. 245703, 2017.

[10] Q. Li *et al.*, "Fluorination of Graphene Enhances Friction Due to Increased Corrugation," *Nano Lett.*, vol. 14, no. 9, pp. 5212–5217, Sep. 2014.

[11] J.-H. Ko *et al.*, "Nanotribological Properties of Fluorinated, Hydrogenated, and Oxidized Graphenes," *Tribol. Lett.*, vol. 50, no. 2, pp. 137–144, 2013.

[12] W. Zhang *et al.*, "Soluble, Exfoliated Two-Dimensional Nanosheets as Excellent Aqueous Lubricants," *ACS Appl. Mater. Interfaces*, vol. 8, no. 47, pp. 32440–32449, Nov. 2016.

[13] X. Han, H. Yong, and D. Sun, "Tuning Tribological Performance of Layered Zirconium Phosphate Nanoplatelets in Oil by Surface and Interlayer Modifications," *Nanoscale Res. Lett.*, vol. 12, no. 1, p. 542, 2017.

[14] F. Jiang, H. Sun, L. Chen, F. Lei, and D. Sun, "Dispersion-tribological property relationship in mineral oils containing 2D layered α-zirconium phosphate nanoplatelets," *Friction*, vol. 8, no. 4, pp. 695–707, 2020.

[15] V. Khare *et al.*, "Graphene–Ionic Liquid Based Hybrid Nanomaterials as Novel Lubricant for Low Friction and Wear," *ACS Appl. Mater. Interfaces*, vol. 5, no. 10, pp. 4063–4075, May 2013.

[16] A. Berardo, G. Costagliola, S. Ghio, M. Boscardin, F. Bosia, and N. M. Pugno, "An experimental-numerical study of the adhesive static and dynamic friction of micro-patterned soft polymer surfaces," *Mater. Des.*, vol. 181, p. 107930, 2019.

[17] G. Costagliola, F. Bosia, and N. M. Pugno, "Tuning friction with composite hierarchical surfaces," *Tribol. Int.*, vol. 115, pp. 261–267, 2017.

[18] G. Costagliola, F. Bosia, and N. M. Pugno, "Static and dynamic friction of hierarchical surfaces," *Phys. Rev. E*, vol. 94, no. 6, p. 63003, Dec. 2016.

[19] A. Niguès, A. Siria, P. Vincent, P. Poncharal, and L. Bocquet, "Ultrahigh interlayer friction in multiwalled boron nitride nanotubes," *Nat. Mater.*, vol. 13, no. 7, pp. 688–693, 2014.

[20] Y. Dong, A. Vadakkepatt, and A. Martini, "Analytical Models for Atomic Friction," *Tribol. Lett.*, vol. 44, no. 3, p. 367, 2011.

[21] L. Liu and M. Eriten, "Frictional Energy Dissipation in Wavy Surfaces," *J. Appl. Mech.*, vol. 83, no. 12, Sep. 2016.

[22] M. Paradinas *et al.*, "Tuning the local frictional and electrostatic responses of nanostructured SrTiO3—surfaces by self-assembled molecular monolayers," *Phys. Chem. Chem. Phys.*, vol. 12, no. 17, pp. 4452–4458, 2010.

[23] J. Zhang *et al.*, "Reduction of interlayer friction between bilayer hexagonal boron nitride nanosheets induced by electron redistribution," *J. Appl. Phys.*, vol. 126, no. 3, p. 35104, Jul. 2019.

[24] P. Gajurel, M. Kim, Q. Wang, W. Dai, H. Liu, and C. Cen, "Vacancy-Controlled Contact Friction in Graphene," *Adv. Funct. Mater.*, vol. 27, no. 47, p. 1702832, Dec. 2017.

[25] J. Kashyap, S. Nagesh, K. Narayan, and P. K. Pattnaik, "Design and simulation of a novel 3D MEMS fabrication/micro cutting facility by thermally actuated MEMS device," in *TENCON 2015 - 2015 IEEE Region 10 Conference*, 2015, pp. 1–4.

[26] T. Filleter *et al.*, "Friction and Dissipation in Epitaxial Graphene Films," *Phys. Rev. Lett.*,

vol. 102, no. 8, p. 86102, Feb. 2009.

[27] X. Zheng et al., "Robust ultra-low-friction state of graphene via moiré superlattice confinement," *Nat. Commun.*, vol. 7, no. 1, p. 13204, 2016.

[28] Y. Zeng, F. He, Q. Wang, X. Yan, and G. Xie, "Friction and wear behaviors of molybdenum disulfide nanosheets under normal electric field," *Appl. Surf. Sci.*, vol. 455, pp. 527–532, 2018.

[29] L.-F. Wang, T.-B. Ma, Y.-Z. Hu, Q. Zheng, H. Wang, and J. Luo, "Superlubricity of two-dimensional fluorographene/MoS2heterostructure: a first-principles study," *Nanotechnology*, vol. 25, no. 38, p. 385701, 2014.

[30] L. Gao et al., "Origin of the moiré superlattice scale lateral force modulation of graphene on a transition metal substrate," *Nanoscale*, vol. 10, no. 22, pp. 10576–10583, 2018.

[31] R. Shi et al., "Moiré superlattice-level stick-slip instability originated from geometrically corrugated graphene on a strongly interacting substrate," *2D Mater.*, vol. 4, no. 2, p. 25079, 2017.

[32] J. Wang, A. Tiwari, Y. Huang, Y. Jia, and B. N. J. Persson, "Dependency of Sliding Friction for Two Dimensional Systems on Electronegativity," Jun. 2020.

[33] M. Li, J. Shi, L. Liu, P. Yu, N. Xi, and Y. Wang, "Experimental study and modeling of atomic-scale friction in zigzag and armchair lattice orientations of MoS2," *Sci. Technol. Adv. Mater.*, vol. 17, no. 1, pp. 189–199, Dec. 2016.

[34] C. Wang, H. Li, Y. Zhang, Q. Sun, and Y. Jia, "Effect of strain on atomic-scale friction in layered MoS2," *Tribol. Int.*, vol. 77, pp. 211–217, 2014.

[35] Z. Pang, J. Wan, A. Lu, J. Dai, L. Hu, and T. Li, "Giant tunability of interlayer friction in graphite via ion intercalation," *Extrem. Mech. Lett.*, vol. 35, p. 100616, 2020.

[36] O. Acikgoz, A. Yanilmaz, O. E. Dagdeviren, C. Çelebi, and M. Z. Baykara, "Inverse Layer Dependence of Friction on Chemically Doped MoS_{2}," Jul. 2020.

[37] H. Terada, H. Imai, and Y. Oaki, "Visualization and Quantitative Detection of Friction Force by Self-Organized Organic Layered Composites," *Adv. Mater.*, vol. 30, no. 27, p. 1801121, Jul. 2018.

[38] S. E, X. Ye, Z. Zhu, W. Lu, C. Li, and Y. Yao, "Tuning the structures of boron nitride nanosheets by template synthesis and their application as lubrication additives in water," *Appl. Surf. Sci.*, vol. 479, pp. 119–127, 2019.

[39] Z. Cheng, G. Zhang, B. Zhang, F. Ma, and Z. Lu, "Tuning the electronic structure of hexagonal boron nitride by carbon atom modification: a feasible strategy to reduce sliding friction," *Mater. Res. Express*, vol. 6, no. 3, p. 36306, 2018.

[40] P. G. Rodriguez et al., "Tuning the Structure and Ionic Interactions in a Thermochemically Stable Hybrid Layered Titanate-Based Nanocomposite for High Temperature Solid Lubrication," *Adv. Mater. Interfaces*, vol. 4, no. 14, p. 1700047, Jul. 2017.

[41] P. L. Dickrell et al., "Tunable friction behavior of oriented carbon nanotube films," *Tribol. Lett.*, vol. 24, no. 1, pp. 85–90, 2006.

[42] T. Arif, G. Colas, and T. Filleter, "Effect of Humidity and Water Intercalation on the Tribological Behavior of Graphene and Graphene Oxide," *ACS Appl. Mater. Interfaces*, vol. 10, no. 26, pp. 22537–22544, Jul. 2018.

[43] Y. Gongyang et al., "Temperature and velocity dependent friction of a microscale graphite-DLC heterostructure," *Friction*, vol. 8, no. 2, pp. 462–470, 2020.

[44] E. Dollekamp, P. Bampoulis, M. H. Siekman, E. S. Kooij, and H. J. W. Zandvliet, "Tuning the Friction of Graphene on Mica by Alcohol Intercalation," *Langmuir*, vol. 35, no. 14, pp. 4886–4892, Apr. 2019.

[45] Z. B. Fredricks, K. M. Stevens, S. G. Kenny, B. Acharya, and J. Krim, "Tuning Nanoscale Friction by Applying Weak Magnetic Fields to Reorient Adsorbed Oxygen Molecules," *Condensed Matter*, vol. 4, no. 1. 2019.

[46] E. Strelcov, R. Kumar, V. Bocharova, B. G. Sumpter, A. Tselev, and S. V Kalinin, "Nanoscale Lubrication of Ionic Surfaces Controlled via a Strong Electric Field," *Sci. Rep.*, vol. 5, no. 1, p. 8049, 2015.

[47] J. Kashyap and D. Datta, "Surface Corrugations and Layer Thickness Dependent Frictional Behavior of $MoS_2$ – A Computational Study," *ECS Meet. Abstr.*, vol. MA2021-02, no. 49, p. 1469, 2021.

[48] J. Kashyap, K. Ghatak, and D. Datta, "Characterizing the Morphology of the Different Grown Homo/Hetero TMD Structures By Controlling Parameters – a Multiscale Computational Approach," *ECS Meet. Abstr.*, vol. MA2019-01, no. 12, p. 806, 2019.

[49] K. Ghatak, D. Datta, J. Kashyap, and K. G. Team, "Growth Physics of MoS2 Layer on the MoS2 Surface: A Monte Carlo Approach," in *APS March Meeting Abstracts*, 2019, vol. 2019, p. F13.009.

[50] X. Li, J. Yin, J. Zhou, and W. Guo, "Large area hexagonal boron nitride monolayer as efficient atomically thick insulating coating against friction and oxidation," *Nanotechnology*, vol. 25, no. 10, p. 105701, 2014.

[51] K. Pivnic, O. Y. Fajardo, F. Bresme, A. A. Kornyshev, and M. Urbakh, "Mechanisms of Electrotunable Friction in Friction Force Microscopy Experiments with Ionic Liquids," *J. Phys. Chem. C*, vol. 122, no. 9, pp. 5004–5012, Mar. 2018.

[52] H. Lee *et al.*, "Nanoscale Friction on Confined Water Layers Intercalated between MoS2 Flakes and Silica," *J. Phys. Chem. C*, vol. 123, no. 14, pp. 8827–8835, Apr. 2019.

[53] X. Shi, S. Liu, Y. Sun, J. Liang, and Y. Chen, "Lowering Internal Friction of 0D–1D–2D Ternary Nanocomposite-Based Strain Sensor by Fullerene to Boost the Sensing Performance," *Adv. Funct. Mater.*, vol. 28, no. 22, p. 1800850, May 2018.

[54] Y. Wu, M. Cai, X. Pei, Y. Liang, and F. Zhou, "Switching Friction with Thermal- Responsive Gels," *Macromol. Rapid Commun.*, vol. 34, no. 22, pp. 1785–1790, Nov. 2013.

[55] Y. Pang *et al.*, "Tribotronic Enhanced Photoresponsivity of a MoS2 Phototransistor," *Adv. Sci.*, vol. 3, no. 6, p. 1500419, Jun. 2016.

[56] F. Xue *et al.*, "MoS2 Tribotronic Transistor for Smart Tactile Switch," *Adv. Funct. Mater.*, vol. 26, no. 13, pp. 2104–2109, Apr. 2016.

[57] G. S. Jung, S. Wang, Z. Qin, F. J. Martin-Martinez, J. H. Warner, and M. J. Buehler, "Interlocking Friction Governs the Mechanical Fracture of Bilayer MoS2," *ACS Nano*, vol. 12, no. 4, pp. 3600–3608, Apr. 2018.

[58] Z. Cui, G. Xie, F. He, W. Wang, D. Guo, and W. Wang, "Atomic-Scale Friction of Black Phosphorus: Effect of Thickness and Anisotropic Behavior," *Adv. Mater. Interfaces*, vol. 4, no. 23, p. 1700998, Dec. 2017.

[59] S. Cahangirov, C. Ataca, M. Topsakal, H. Sahin, and S. Ciraci, "Frictional Figures of Merit for Single Layered Nanostructures," *Phys. Rev. Lett.*, vol. 108, no. 12, p. 126103, Mar. 2012.

[60] A. I. Volokitin and B. N. J. Persson, "Near-field radiative heat transfer and noncontact friction," *Rev. Mod. Phys.*, vol. 79, no. 4, pp. 1291–1329, Oct. 2007.

[61] F. Giacco, M. P. Ciamarra, L. Saggese, L. de Arcangelis, and E. Lippiello, "Non-monotonic dependence of the friction coefficient on heterogeneous stiffness," *Sci. Rep.*, vol. 4, no. 1, p. 6772, 2014.

[62] P. Manimunda *et al.*, "Nanoscale deformation and friction characteristics of atomically thin WSe 2 and heterostructure using nanoscratch and Raman spectroscopy," *2D Mater.*, vol. 4, no. 4, p. 45005, 2017.

[63] L. Bai, B. Liu, N. Srikanth, Y. Tian, and K. Zhou, "Nano-friction behavior of phosphorene," *Nanotechnology*, vol. 28, no. 35, p. 355704, 2017.

[64] X. Zhou *et al.*, "Influence of elastic property on the friction between atomic force microscope tips and 2D materials," *Nanotechnology*, vol. 31, no. 28, p. 285710, 2020.

[65] J. Kashyap and D. Datta, "Drug repurposing for SARS-CoV-2: a high-throughput molecular docking, molecular dynamics, machine learning, and DFT study," *J. Mater. Sci.*, 2022.

[66] W. Tang, E. Sanville, and G. Henkelman, "A grid-based Bader analysis algorithm without lattice bias," *J. Phys. Condens. Matter*, vol. 21, no. 8, p. 84204, 2009.

[67] Y. Hinuma, G. Pizzi, Y. Kumagai, F. Oba, and I. Tanaka, "Band structure diagram paths based on crystallography," *Comput. Mater. Sci.*, vol. 128, pp. 140–184, 2017.

[68] A. Gulans *et al.*, "exciting: a full-potential all-electron package implementing density-functional theory and many-body perturbation theory," *J. Phys. Condens. Matter*, vol. 26, no. 36, p. 363202, 2014.

[69] J. F. Archard and T. E. Allibone, "Elastic deformation and the laws of friction," *Proc. R. Soc. London. Ser. A. Math. Phys. Sci.*, vol. 243, no. 1233, pp. 190–205, Dec. 1957.

**Supplementary Information: Engineering frictional characteristics of MoS$_2$ structure by tuning thickness and morphology- An atomic, electronic structure, and exciton analysis**

Jatin Kashyap*[1], Joseph Torsiello [2][4], Yoshiki Kakehi [3][5], Dibakar Datta[1]

[1] Department of Mechanical and Industrial Engineering, New Jersey Institute of Technology Newark, NJ 07103, USA
[2] Department of Physics, New Jersey Institute of Technology, Newark, New Jersey 07102, USA
[3] Bergen County Technical Schools, Teterboro, NJ 07608, USA
[4] Department of Physics, Temple University, Philadelphia, PA 19122, USA
[5] College of Computing, Georgia Institute of Technology, Atlanta, GA 30332, USA

*Corresponding authors
Jatin Kashyap, Email: jk435@njit.edu

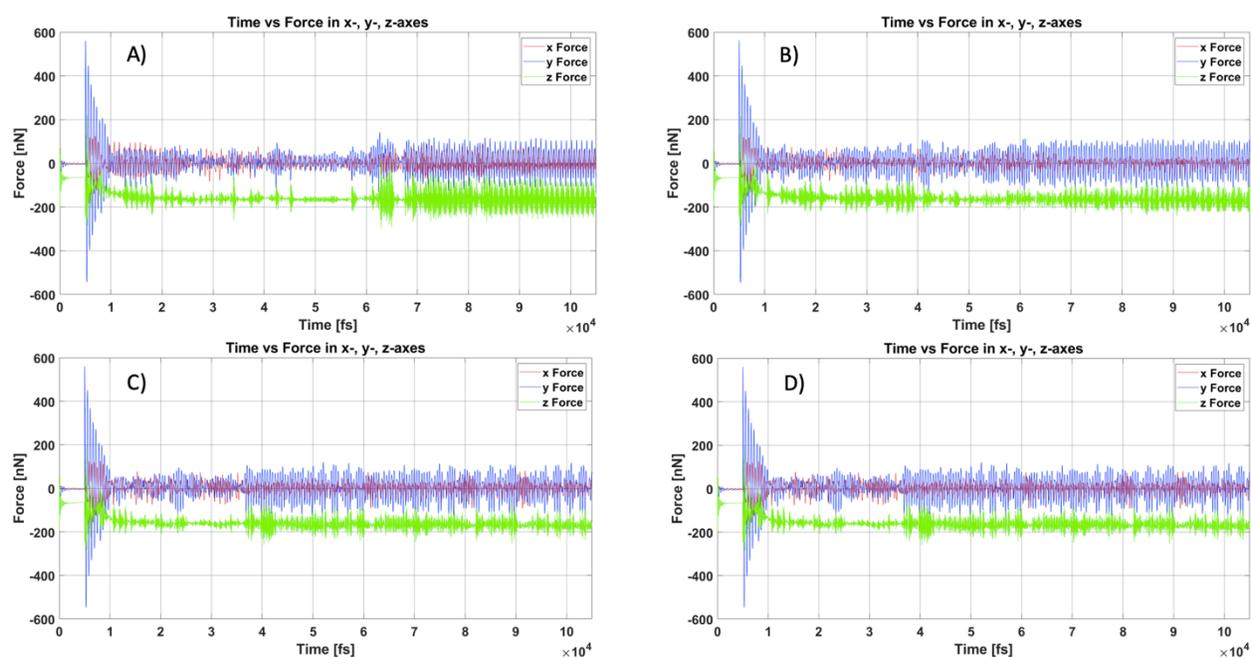

*Figure S11 Force plots for radius subcase. A), B), C) & D) shows the plots for 12 Å, 16 Å, 20 Å, & 24Å respectively.*

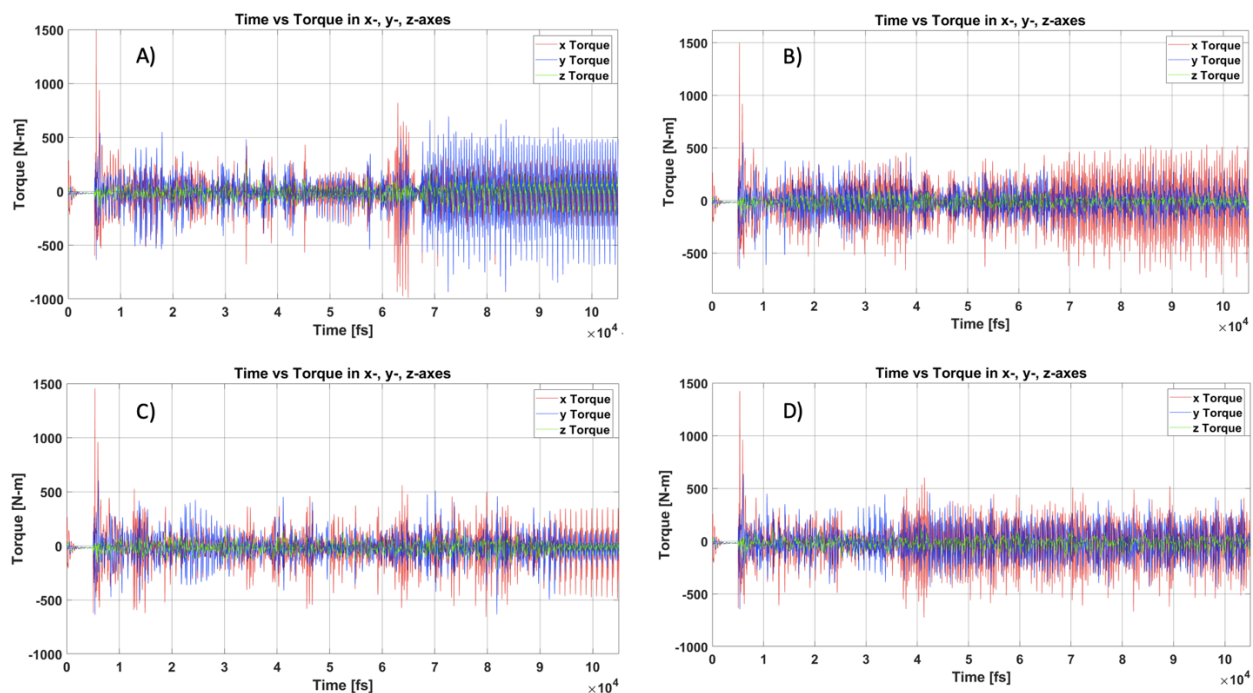

*Figure S12 Torque plots for radius subcase. A), B), C) & D) shows the plots for 12 Å, 16 Å, 20 Å, & 24Å respectively.*

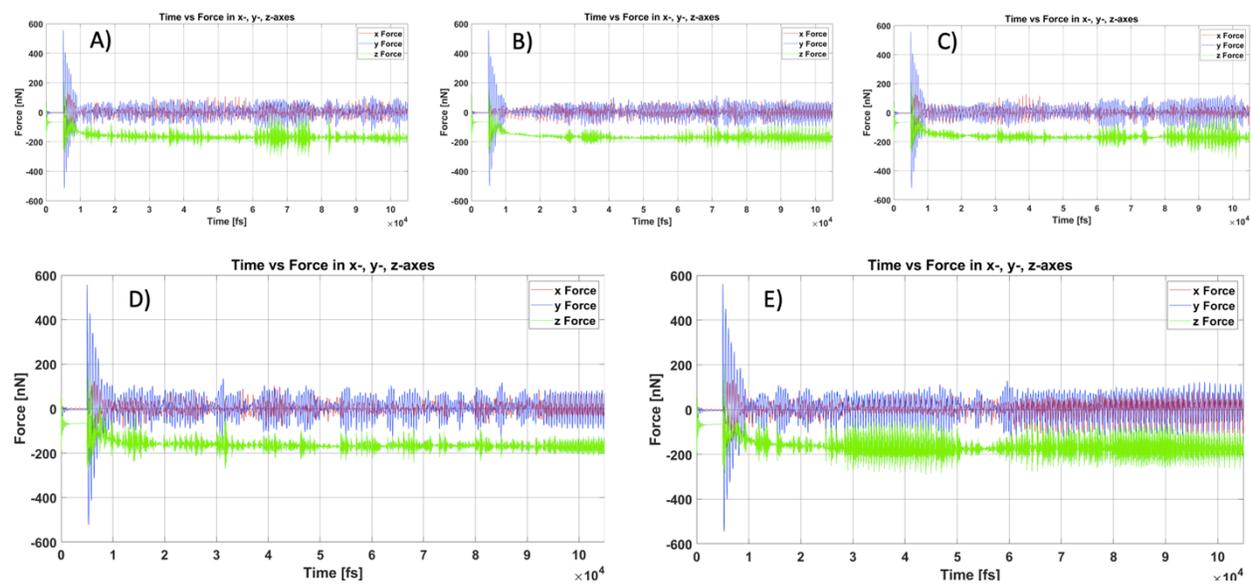

*Figure S13 Force plots for indents' pattern subcase. A), B), C), D) & E) shows the plots for Θ=0°, 25°, 30°, 35°, 45°, 60° respectively.*

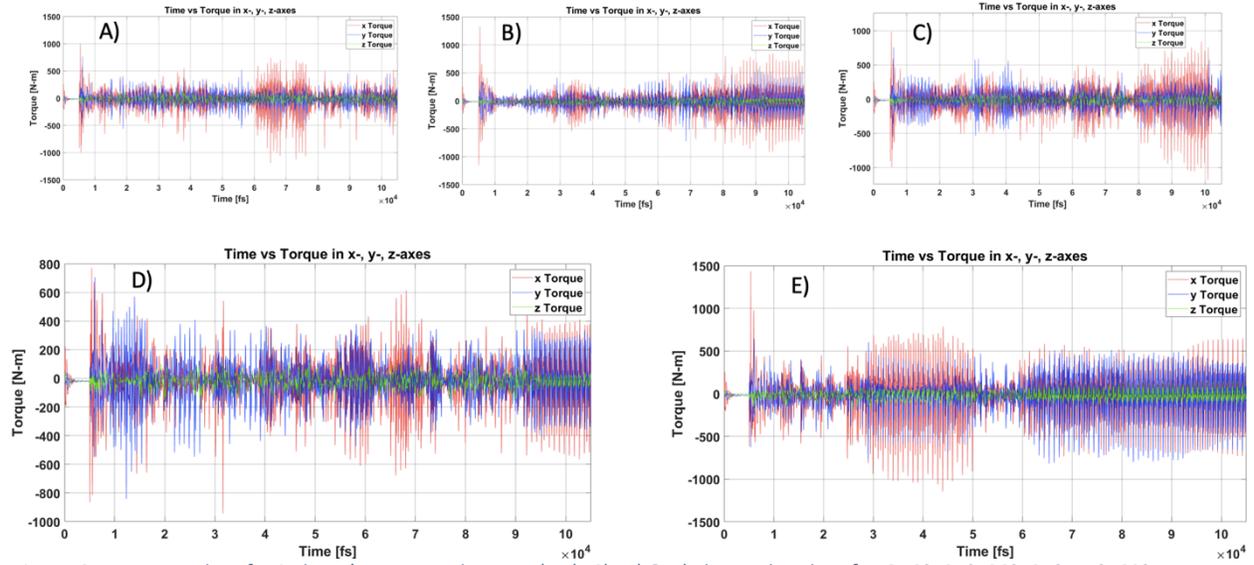

Figure S14 Torque plots for indents' pattern subcase. A), B), C), D) & E) shows the plots for Θ=0°, 25°, 30°, 35°, 45°, 60° respectively.

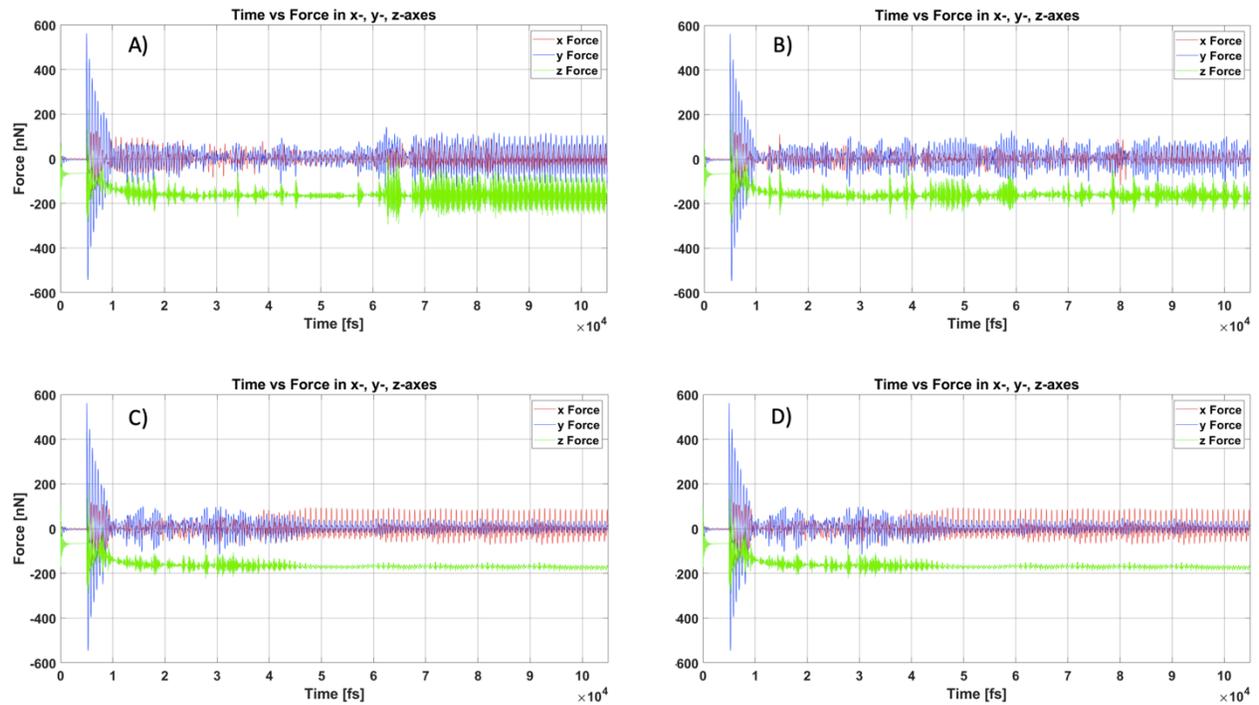

Figure S15 Force plots for number of layers subcase. A), B), C) & D) shows the plots for 1, 2, 3, & 4 layers respectively.

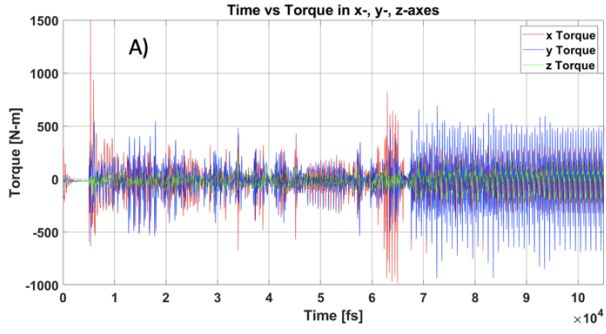
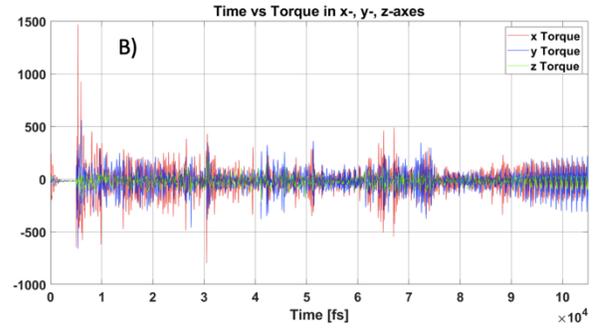
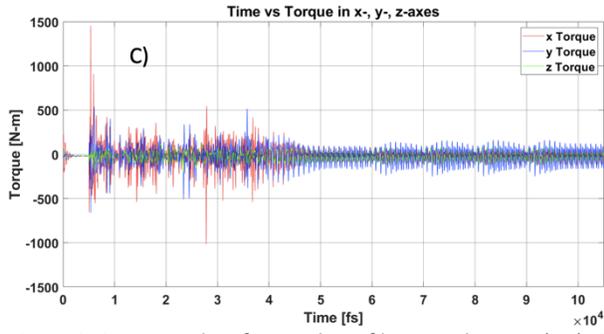
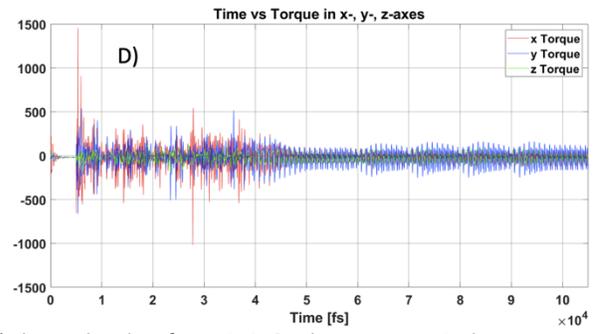

*Figure S16 Torque plots for number of layers subcase. A), B), C) & D) shows the plots for 1, 2, 3, & 4 layers respectively.*

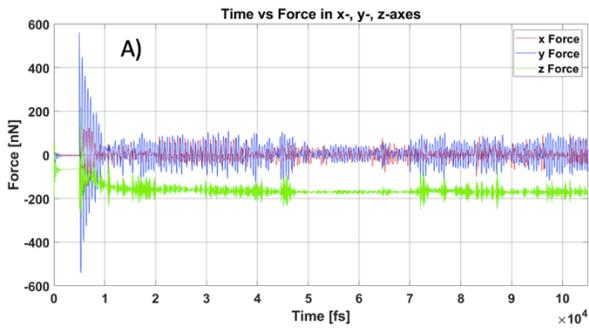
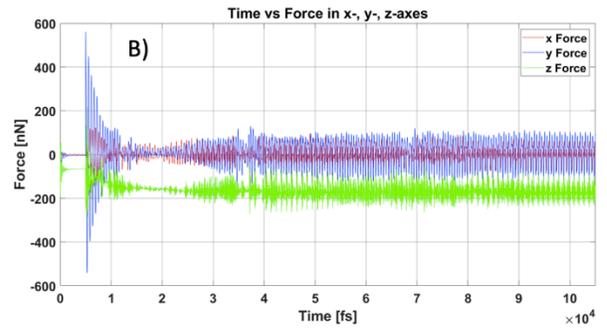
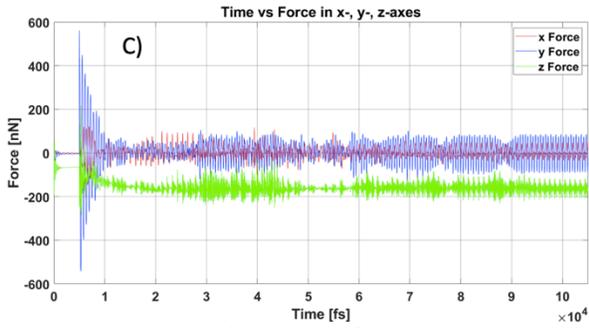
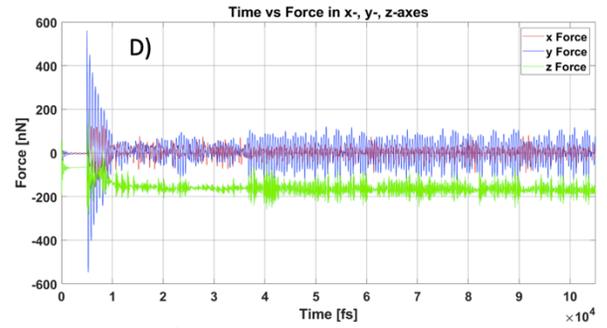

*Figure S17 Force plots for number of indents subcase. A), B), C) & D) shows the plots for 2, 4, 6, & 8 indents respectively.*

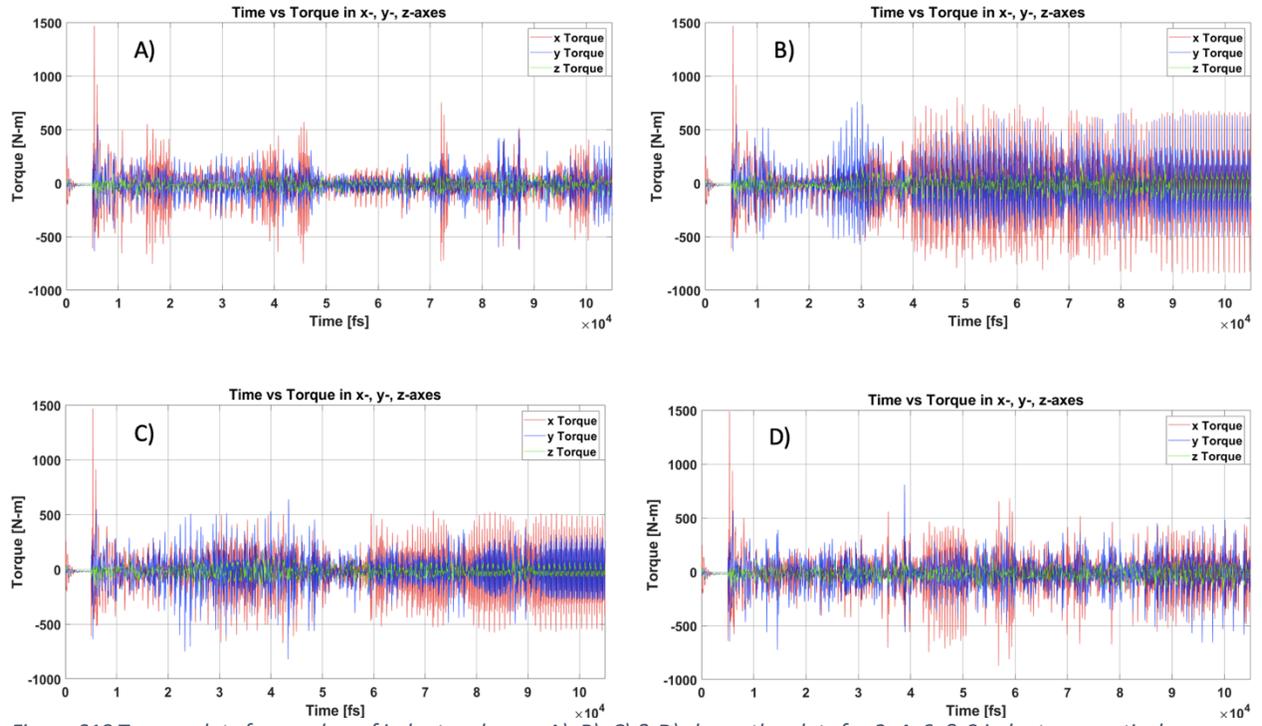
Figure S18 Torque plots for number of indents subcase. A), B), C) & D) shows the plots for 2, 4, 6, & 8 indents respectively.

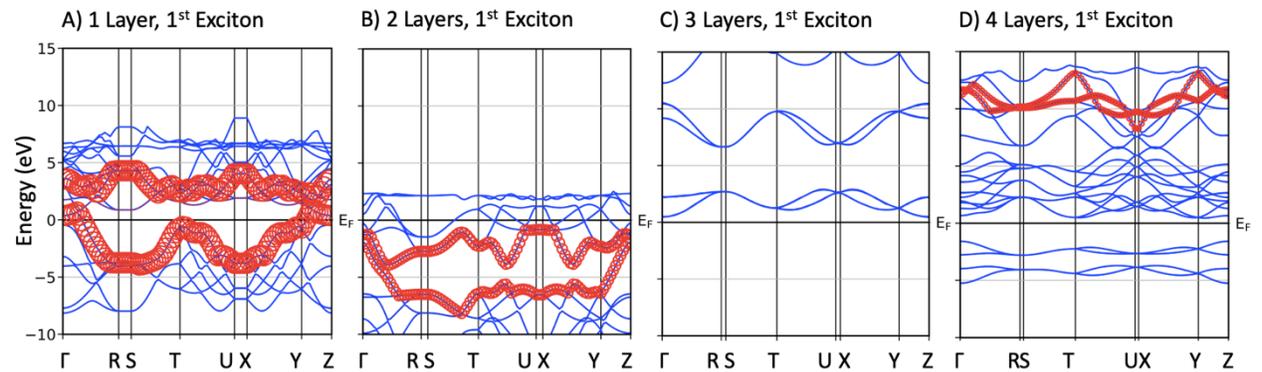
Figure S19 Fatbands structures of 1st excitons. A), B), C) & D) are showing electron-hole couples for 1, 2, 3, & 4 layers cases respectively.

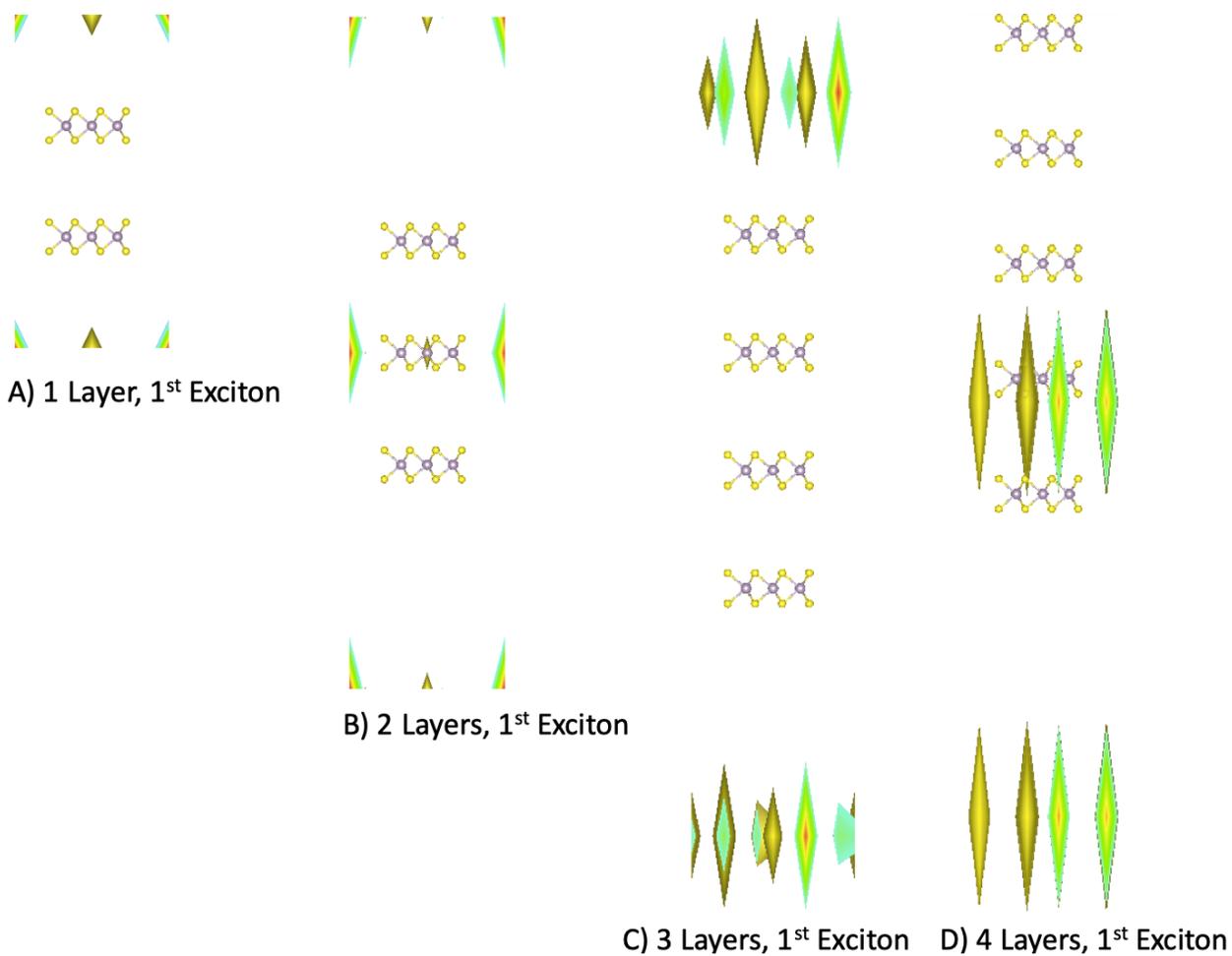

Figure S20 Real space visualization of 1st excitons. A), B), C) & D) are showing electron-hole couples' clouds for 1, 2, 3, & 4 layers cases respectively.